\documentclass[journal=jpccck,  manuscript=article,layout=twocolumn]{achemso}

\usepackage[utf8]{inputenc}
\usepackage{graphicx}
\usepackage{float}

\usepackage{amsfonts}
\usepackage{multirow}
\usepackage{setspace}
\usepackage{achemso}
\usepackage{tabularx}
\usepackage{tikz}
\usepackage{url}
\usepackage{amssymb}
\usepackage{amsmath}
\usepackage[labelformat = empty,position=top]{subcaption} 
\graphicspath{ {./Figures/} }
\usepackage{enumitem}
\usepackage{changes}

\definecolor{columbiablue}{rgb}{0.61, 0.87, 1.0}
\definecolor{celadon}{rgb}{0.67, 0.88, 0.69}
\definecolor{aquamarine}{rgb}{0.5, 1.0, 0.83}
\definecolor{electriclime}{rgb}{0.8, 1.0, 0.0}
\usepackage{todonotes}

\usepackage{soul}
\sethlcolor{white!0}
\usepackage{booktabs}
\usepackage{adjustbox}

\author{Fabian L. Thiemann}
\affiliation[UCL2]{Thomas Young Centre, London Centre for Nanotechnology, and
Department of Physics and Astronomy, University College London,
Gower Street, London, WC1E 6BT, United Kingdom}
\alsoaffiliation[Imperial2]{Department of Chemical Engineering, Imperial College London, South Kensington Campus, London SW7 2AZ, United Kingdom}

\author{Patrick Rowe}
\affiliation[UCL2]{Thomas Young Centre, London Centre for Nanotechnology, and
Department of Physics and Astronomy, University College London,
Gower Street, London, WC1E 6BT, United Kingdom}

\author{Erich A. Müller}
\affiliation[Imperial2]{Department of Chemical Engineering, Imperial College London, South Kensington Campus, London SW7 2AZ, United Kingdom}

\author{Angelos Michaelides}
\affiliation[UCL2]{Thomas Young Centre, London Centre for Nanotechnology, and
Department of Physics and Astronomy, University College London,
Gower Street, London, WC1E 6BT, United Kingdom}
\email{angelos.michaelides@ucl.ac.uk}

 \title{Machine Learning Potential for Hexagonal Boron Nitride Applied to Thermally and Mechanically Induced Rippling}


\begin{document}

\begin{tocentry}
\vspace{0.25cm}
    \includegraphics[width=\textwidth]{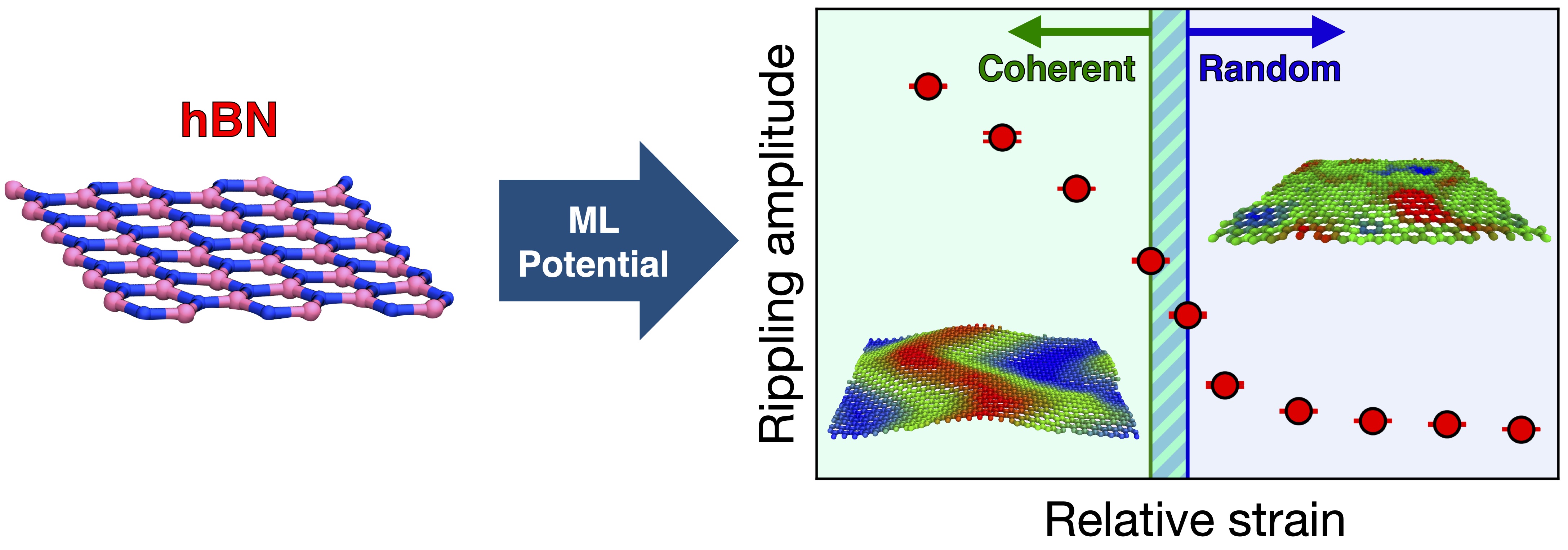}
\end{tocentry}

\begin{abstract}
We introduce an interatomic potential for hexagonal boron nitride (hBN) based on the Gaussian approximation potential (GAP) machine learning methodology. The potential is based on a training set of configurations collected from density functional theory (DFT) simulations and is capable of treating bulk and multilayer hBN as well as nanotubes of arbitrary chirality. The developed force field faithfully reproduces the potential energy surface predicted by DFT while improving the efficiency by several orders of magnitude. We test our potential by comparing formation energies, geometrical properties, phonon dispersion spectra and mechanical properties with respect to benchmark DFT calculations and experiments. In addition, we use our model and a recently developed graphene-GAP to analyse and compare thermally and mechanically induced rippling in large scale two-dimensional (2D) hBN and graphene. \hl{Both materials show almost identical scaling behaviour with an exponent of $\eta \approx 0.85$ for the height fluctuations agreeing well with the theory of flexible membranes.} Based on its lower resistance to bending, however, hBN experiences slightly larger out-of-plane deviations both at zero and finite applied external strain. Upon compression a phase transition from incoherent ripple motion to soliton-ripples is observed for both materials. Our potential is freely available online at [\textcolor{blue}{\url{http://www.libatoms.org}}].
\end{abstract}

\section{Introduction}

Hexagonal boron nitride (hBN), though less well-known than its isostructural analogue, graphene, is no less fascinating from the perspective of its properties and potential applications.
The insulating character\cite{Watanabe2004}, resistance to oxidation \cite{Chen2004}, high thermal conductivity, \cite{Pakdel2014} and mechanical strength \cite{Geim2013} of hBN have been exploited in various fields ranging from electronic devices \cite{Britnell2012,Xia2014a}, water purification \cite{Lei2013}, and industrial chemistry \cite{Zhi2011,Taha-Tijerina2012,Krishnam2016,Zhao2019} to 
biomedical science and engineering \cite{Ciofani2013,Weng2014,Weng2016}.
Yet, certain aspects of the behaviour of low-dimensional hBN remain to be rationalised and explored further. For instance, the origin of experimentally observed differences in friction \cite{Secchi2016} and surface charge \cite{Siria2013,Grosjean2016} between graphene and hBN surfaces in contact with water requires a deeper understanding on the microscopic level \cite{Secchi2016a} to exploit the full potential of hBN for desalination processes and osmotic power generation \cite{Siria2017}. 
Similarly, the impact of intrinsic ripples on layered hBN has not received as much attention as for graphene \cite{Meyer2007,Fasolino2007}, where these out-of-plane deviations have a significant impact on properties such as graphene's bandgap \cite{Guinea2008,Guinea2008a,VazquezDeParga2008} or reactivity \cite{Boukhvalov2009}. 
Assuming analogous behaviour for two-dimensional (2D) hBN the ability to predict, and potentially adjust the ripple texture through strain would be of value to future applications \cite{Bao2009}.\\

Molecular dynamics (MD) simulations offer a computational approach to gain insight into nanoscale properties of materials.
A faithful representation of the potential energy surface (PES) is a crucial requirement for reliable simulations, in particular an accurate description of the phonon dispersion spectrum which determines a variety of mechanical and thermodynamic properties as well as diffusion processes across low-dimensional materials.
Due to the strong coupling between the adsorbate's motion and the phonon modes of the solid \cite{Ma2015,Ghoufi2016,Cruz-Chu2017,Marbach2018}, potential inaccuracies in the predicted lattice vibrations are directly propagated to dynamical and interfacial properties.
Various force fields for hBN are available, including Tersoff-type potentials \cite{Tersoff1986,Tersoff1988,Tersoff1989,Brenner1990,Albe1997,Sekkal1998,Matsunaga2000,Matsunaga2001,Sevik2011,Kinaci2012,Los2017}, reactive force fields \cite{VanDuin2001,Chenoweth2008,Weismiller2010,Paupitz2014,Liu2017}, and potentials fitted to density functional theory (DFT) \cite{Rajan2018}. Despite the valuable and excellent work on thermal and mechanical properties of hBN allotropes \cite{Mortazavi2012,Liao2012,AnoopKrishnan2014,Le2014} based on these established models and notwithstanding their computational efficiency, none of them predicts the vibrational properties in good agreement with experiments as we learned in the course of this work. Thus, these models cannot be expected to give reliable results for our systems of interest.\\

Incorporating electronic structure based methods and performing \textit{ab initio} molecular dynamics (AIMD) represents an alternative approach which often offers a significantly more accurate description of the PES and the phonon dispersion spectrum. 
DFT driven AIMD has been employed to study hBN in various contexts including proton transfer through a hBN sheet \cite{Hu2014}, interfacial behaviour with water \cite{Tocci2014a}, and the impact of strain on the vibrational properties of single and multiple layers of hBN \cite{Androulidakis2018}.
These methods, however, come with a high computational cost that severely restricts the accessible system sizes and time scales.
Finite-size effects remain a practical issue of AIMD simulations and can compromise the reliability of calculated transport properties for confined fluids \cite{Simonnin2017}.
This is also true for ripples where the amplitude of the out-of-plane deviations is partially determined by the dimensions of the 2D layer \cite{Fasolino2007,Los2009}.\\

The rise of machine learning (ML) methodologies and their fruitful application in the development of interatomic potentials has provided a pathway for achieving an accuracy close to that of their \textit{ab initio} reference while lowering the computational costs by several orders of magnitude\hl{{ \cite{Rupp2012,Behler2016,Csanyi2017,Botu2017,Li2017,Kitchin2018,Goldsmith2018}}}. These potentials are based on a transformation of atomic coordinates using high dimensional descriptors \cite{Behler2016} which serve as input for the ML algorithm to establish a structure-energy mapping. In order to actually ``learn'' this relationship, a large database of configurations and related \textit{ab initio} energies and forces is required. In recent years, many ML-based potentials have been developed for a significant number of materials spanning a variety of algorithms including Gaussian kernel regression, artificial neural networks and permutationally invariant polynomials \cite{Behler2007, Bartok2010,Li2015,Chmiela2017a,Nguyen2018}. 
A machine learning potential based on the Gaussian Approximation Potential (GAP) \cite{Bartok2010,Bartok2015} framework was recently presented by some of us for hBN’s isostructural analogue, graphene \cite{Rowe2018}. This model achieves a very good agreement with experiments throughout a large range of properties including the thermal expansion and phonon dispersion spectrum. GAP models also have been proposed to study amorphous and crystalline carbon \cite{Deringer2017,Rowe2020b}, tungsten \cite{Szlachta2014}, silicon \cite{Bartok2010}, as well as hybrid perovskites \cite{Jinnouchi2019}. \\

In this work we introduce a GAP for hBN which is able to treat the bulk phase, isolated and multilayered sheets of hBN, as well as nanotubes of arbitrary chirality.
We evaluate the performance of our model by comparing against DFT as well as frequently applied force fields for a variety of properties.
In particular, significant improvements for the phonon spectra are obtained with the hBN-GAP compared to established force fields.
We, then, perform large-scale MD simulations using our new potential and the recently published graphene-GAP \cite{Rowe2018} to investigate the behaviour of thermally and mechanically induced rippling in 2D hBN and graphene.
We find an almost identical scaling exponent for both materials, while the amplitude and shape of the height fluctuations depends on the material's properties as well as on the applied strain.
Looking forward, the hBN-GAP will be particularly valuable where an accurate description of the system's vibrations and dynamics is required \hl{and where the desired length and timescales are not accessible with electronic structure methods.}
This includes quantitative studies on diffusion of adsorbates across layered hBN or through hBN nanotubes.\\

The remainder of this paper is structured as follows. Next, we give a brief overview of the GAP methodology as well as a detailed explanation of technical aspects which are crucial to consider in the construction of a ML potential. This is followed by a detailed \hl{evaluation} of the accuracy of the hBN-GAP. The results of the rippling analysis are reported before we draw the conclusions of this study.

\section{\hl{Theory and Computational Details}}

\subsection*{\hl{Construction of the hBN-GAP}}

A detailed derivation of the theory behind the GAP framework is available elsewhere \cite{Bartok2010,Szlachta2014,Bartok2015,Deringer2017,Rowe2018}. For completeness, we will \hl{give} an abridged explanation and focus on the practical issues within the construction. GAP mimics the Born-Oppenheimer PES without treating electrons explicitly by establishing a link between structure and energy based on a transformation of the atomic positions into local environments using so-called descriptors \cite{Bartok2010}. To this end, the target PES is represented by a large database of \textit{ab initio} energies which is then interpolated by using Gaussian kernel regression (GKR). The decomposition into local contributions is an approximation inherent to all interatomic potentials and can vary from atom-pair distances to complex many-body (MB) descriptors \cite{Behler2016}. Independent of their dimensionality, however, descriptors are generally required to be translationally, rotationally, and permutationally invariant to guarantee a one to one mapping of energy to configuration \cite{De2016}.\\

In this work, the PES of hBN is described by a sum of two-body (2B) terms and the high dimensional smooth overlap of atomic positions (SOAP) \cite{Bartok2013} MB descriptor. Within SOAP, each atomic environment is represented by a local neighbour density generated by summing over Gaussians placed on all atoms within a certain cut-off. \hl{This density of the $i$th atom, $\rho_i (\mathbf{r})$, is expanded in a basis set of radial functions $g_n(r)$ and spherical harmonics $Y_{lm}(\mathbf{r})$ as}

\begin{equation}
  \rho_i(\mathbf{r})=  \sum_{\substack{n<n_\mathrm{max} \\ l<l_\mathrm{max} \\ \left|m \right| \leq l }} c_{nlm}^i g_n(r) Y_{lm}(\mathbf{r}) \mbox{ , }
  \label{eq:self_energy}
\end{equation}

\hl{wherein the related coefficients $c^i_{nlm}$ form the so-called SOAP vectors with $n$,$l$, and $m$ being the familiar integers}. The product of two independent SOAP vectors represents the similarity between the related local environments, which can be easily made invariant to rotations. This so-called SOAP kernel is constructed for a certain set of training environments and the corresponding weight of their local contribution to the total energy of the global configurations is determined within the GKR fitting procedure. In addition to the energies, we also train on the atomic forces and virial stresses. A deeper insight and detailed derivation of the extension to use partial derivatives can be found elsewhere   \cite{Bartok2015}. \\

Clearly, an accurate description of the PES is critically dependent on a comprehensive training set comprised of configurations of the desired phase space region. Here, this covers periodic and defect-free sheets and nanotubes of arbitrary chiralities. We generated a structural database containing more than 22,000 configurations. \hl{While most of them were extracted from AIMD trajectories, others originated from trajectories obtained with earlier GAP versions or were based on distinct displacements of atoms. In fact, including the latter configurations at 0
~K is crucial to accurately predict the equilibrium configurations, lattice parameters or elastic constants. The number of atoms was limited to 200 and different thermodynamic conditions were sampled varying the temperature between 0 and 3000 K as well as applying uniaxial strain of up to 5~\% to layered hBN which should also enable the prediction of strained nanotubes. These conditions represent the region of phase space where the hBN-GAP can be safely applied and an accurate interpolation can be expected. } \\

In practice, it is not desirable to use the entire structural database as training set due to the redundant information and the related increase of computational costs. Here a subset of $\approx$ \hl{1600} structures was used. These were selected through a combination of farthest point sampling \cite{De2016,Csanyi2017} and manual structure selection. \hl{While this has been shown to give an accurate representation of the original database, the training set could be further improved by using active learning which adds new configurations based on reliability of the model's prediction rather than the structural diversity in the existing training set {\cite{Smith2018}}.}\\

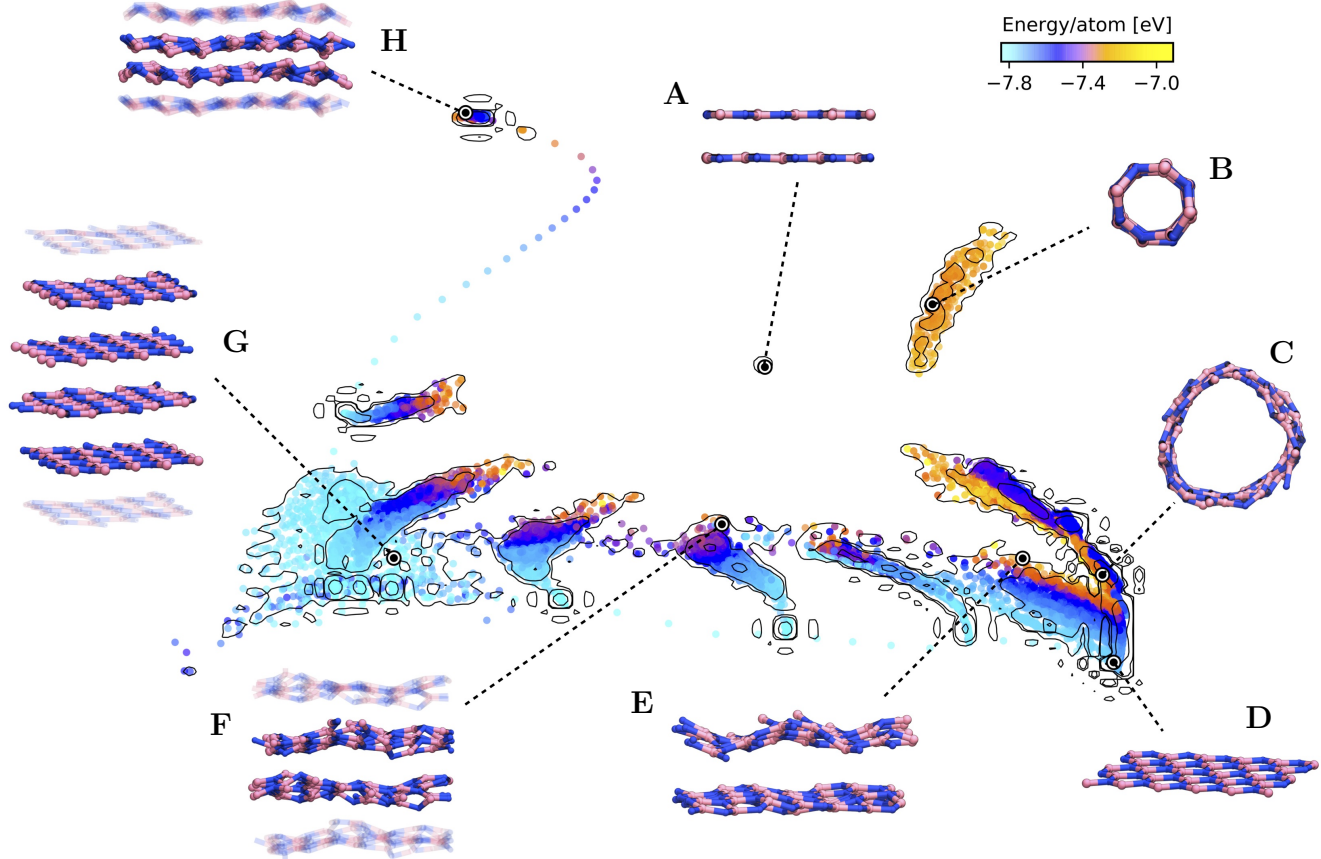
\begin{figure*}[ht!]
    \centering
    \begin{tikzpicture}
    \node (fig) at (0,0)    {\includegraphics[width=\textwidth]{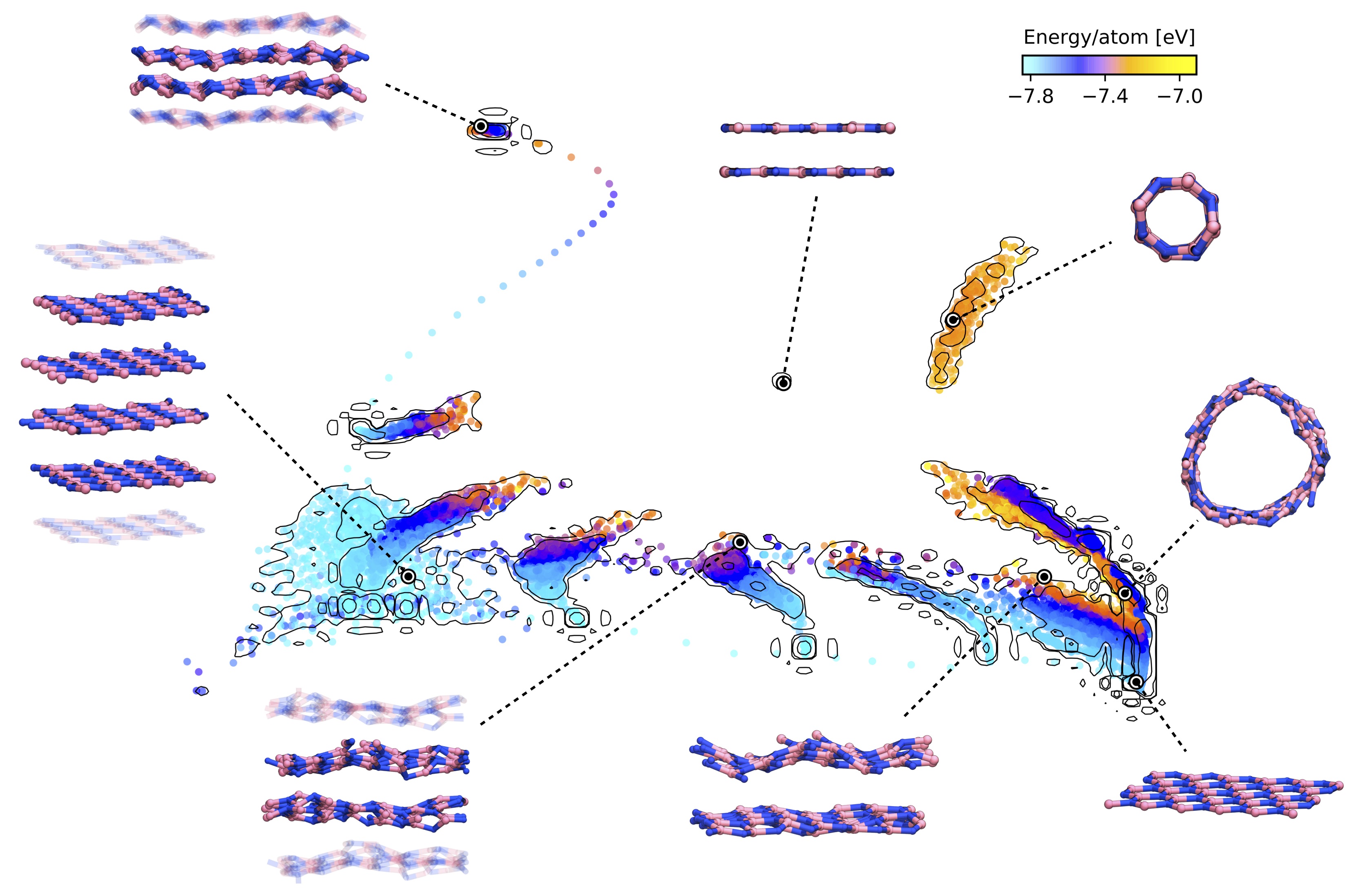}};
    \node at (0.25,4.5) {\textbf{A} };
    \node at (7.5,3.5) {\textbf{B} };
    \node at (8.3,1.05) {\textbf{C} };
    \node at (8.0,-3.8) {\textbf{D} };
    \node at (-0.2,-3.6) {\textbf{E} };
    \node at (-5.8,-3.9) {\textbf{F} };
     \node at (-5.6,1.2) {\textbf{G} };
       \node at (-3.5,5.2) {\textbf{H} };
   \end{tikzpicture}
    \caption {Sketch-map representation of the structural database generated for this work. Each point corresponds to one atomic configuration for which the local SOAP vectors are averaged and the SOAP kernel measures the global similarity between the structures. Short distances between points represent higher similarity while points far apart from each other express a strong structural deviation. The points are coloured according to their respective energy per atom calculated with DFT. Periodic images in z-direction are blurred to visualise bulk structures. (\textbf{A}) Compressed bilayer at 0~K (interlayer spacing $\approx 2.5$~$\mathrm{\AA}$); (\textbf{B}) Nanotube (4,0) at 1000~K; (\textbf{C}) Nanotube (10,0) at 300~K; (\textbf{D}) Monolayer at 30~K;(\textbf{E}) Bilayer at 3000~K; (\textbf{F}) Bulk at 3000~K; (\textbf{G}) Bulk at 300~K; (\textbf{H}) Compressed bulk at 700~K. \label{fig:Sketchmap}}
   
\end{figure*}

In order to obtain a qualitative overview of the complexity of problem we are attempting to fit, we performed a dimensionality reduction using sketch-map \cite{Ceriotti2011,Tribello2012} to visualise our structural database in two dimensional space. The resulting sketch-map, shown in figure \ref{fig:Sketchmap}, clusters configurations based on their global similarity. Similar structures are located closely together while points with a high covariance are separated further apart. The clustering indicates the heterogeneity of the phase space region and provides useful information on the capabilities of the SOAP descriptor. For example, the sketch-map highlights the anticipated similarity between monolayers (Fig. \ref{fig:Sketchmap}D) and nanotubes with large diameter (Fig. \ref{fig:Sketchmap}C) while less stable thermally deformed nanotubes with high curvature are represented as an isolated island (Fig. \ref{fig:Sketchmap}B). Single layer and bulk hBN (Fig. \ref{fig:Sketchmap}G) are separated most far apart given the change of environment induced by the adjacent layers. This dissimilarity distance is bridged by bulk configurations with large interlayer spacing, strong thermal fluctuations (Fig. \ref{fig:Sketchmap}F) and bilayers (Fig. \ref{fig:Sketchmap}E). Interestingly, compressed bilayers (Fig. \ref{fig:Sketchmap}A) are singled out from the large clusters as well as sheared bulk configurations (Fig. \ref{fig:Sketchmap}H). \\

Having selected the training configurations, an appropriate choice of certain model parameters has to be made. Here, we will only discuss those parameters to which the model is most sensitive while a full overview is contained in table \ref{tab:hyperparameters} and a more detailed explanation can be found elsewhere \cite{Szlachta2014,Bartok2015}. Intuitively, cut-off and basis-set expansion of the SOAP vector have a high impact on the performance of the potential. We chose to truncate the spherical harmonic expansion after the eighth order, i.e. $l=8$ and $n=8$, and set the SOAP cut-off to 4.5~$\mathrm{\AA}$ which has been proven to be sufficient to reproduce the binding energy curve of graphite \cite{Rowe2020b}. However, this cut-off for the SOAP descriptor is too short to reproduce the interlayer interaction curve between hBN layers which does not tail off to zero until roughly 10~$\mathrm{\AA}$. Thus, we followed the procedure recently introduced \cite{Rowe2020b} to fit a 2B-based model to the interaction curve using a cut-off of 10~$\mathrm{\AA}$. The predicted energy and forces were then subtracted during the actual fit where we used a shorter 2B descriptor with the same cut-off as SOAP. \\

\begin{table}[h]
    \caption{Model parameters for the hBN-GAP. \hl{Target deviations ($\sigma$) correspond to the range, with values depending on the specific configuration}.}
    \begin{adjustbox}{center} 
        \begin{tabular}{lr}
             \toprule
                2b Descriptor & \\
             \cmidrule{1-1}
                Cut-off $(\mathrm{\AA})$ & 4.5 \\
                Cut-off width $(\mathrm{\AA})$ & 0.5 \\
                $\delta$ &  0.5\\
                Sparse method & Uniform \\
                Sparse points & 50 \\
            \cmidrule{1-1}
                SOAP Descriptor & \\
            \cmidrule{1-1}
                Cut-off $(\mathrm{\AA})$ & 4.5 \\
                Cut-off width $(\mathrm{\AA})$ & 0.5 \\
                $\delta$ & 0.1 \\
                Sparse method & CUR \\
                Sparse points & 5000 \\
                $l_{\mathrm{max}}$ & 8 \\
                $n_{\mathrm{max}}$ & 8 \\
                $\zeta$ & 2 \\
            \cmidrule{1-1}
                Target Deviations  & \\
            \cmidrule{1-1}
                $\sigma_{\mathrm{energy}}$ &\hl{0.0001 - 0.0015}  \\
                $\sigma_{\mathrm{force}}$ & \hl{0.001 - 0.015} \\
                $\sigma_{\mathrm{virial}}$ & \hl{0.005 - 0.05} \\
             \bottomrule
        \end{tabular}
    \end{adjustbox}
    \label{tab:hyperparameters}
\end{table}

In practice, rather than using all local environments of the training set, the covariance matrix is constructed based on $M$ uncorrelated sparse points \cite{Szlachta2014}. Here, we chose 50 points for the 2B descriptor and 5000 for SOAP. The number of sparse points needed depends on the complexity of the target phase space region the model is fitted to. Given the focus on the defect-free hexagonal phase of BN, the selected number of sparse points for SOAP is in line with recent work \cite{Rowe2020b}. Besides these descriptor related parameters, the desirable target deviation, $\sigma$, of the fit to the target properties is a critical ingredient for an accurate model. While values which are too large result in poor agreement with the DFT reference \textit{per se}, too small a chosen value can lead to overfitting depriving the GKR of its interpolation capability. Here, we used configuration-specific tolerances in order to distinguish between competing phases. For example, the PES of bulk hBN is very shallow close to the equilibrium resulting in energy differences of distorted cells of less than 1 meV. Thus, the tolerance was adapted for this kind of system while for high energy configurations larger values were set.

\subsection*{\hl{Electronic structure calculations}}

For each configuration, we conducted tightly converged DFT calculations to obtain the total energy, atomic forces, and virial stresses on which the GAP model was then trained. For all configurations we used the VASP plane-wave DFT code \cite{Kresse1993,Kresse1994,Kresse1996,Kresse1996a}, \hl{using the Perdew-Burke-Ernzerhof (PBE) functional {\cite{Perdew1996}} with the DFT-D3 dispersion correction method {\cite{Grimme2010}} employing the Becke-Johnson damping {\cite{Grimme2011}}, an energy cut-off of 1100 eV, a Gaussian smearing of 0.05 eV, and projector augmented wave pseudopotentials {\cite{Blochl1994,Kresse1999}}. We chose PBE+D3 as it accurately predicts the out-of-plane lattice parameter of bulk hBN with respect to experiments. A comparison to other functionals is included in the supplementary information. The seemingly high plane-wave cut-off is based on convergence tests for the elastic properties of bulk hBN.} The reciprocal lattice was sampled in periodic directions with a maximum distance between the \textit{k}-points of 0.02 $\mathrm{\AA^{-1}}$ whereby the grid was either centered at the zone centre in case of hexagonal cells or Monkhorst-Pack based for orthorombic cells. For ionic and cell relaxations, cell shape and atomic positions were optimised independently until all forces were below $10^{-3}$ eV $\mathrm{\AA}^{-1}$ and the total energy \hl{difference} was converged to less than $10^{-6}$ eV. These settings were consistently used in this work if not stated otherwise.\\

\subsubsection*{\hl{Molecular Dynamics Simulations}}

\hl{The simulations of rippled graphene and hBN were} performed in LAMMPS \cite{Plimpton1997} using a hexagonal simulation box comprising 7200 atoms applying a timestep of 1 fs. 
The unstrained sheets span a size of $15\times15$ nm varying slightly according to the respective bond length.
While this scale is beyond the usually accessible system size for DFT-based MD, GAP models can easily cope with these dimensions. 
First, we sampled the equilibrium lattice parameter of both materials at $300$~K  and zero external pressure in the isothermal-isobaric ensemble.
A Nos\'e-Hoover chain thermostat and barostat was applied to ensure the target temperature and pressure. 
Then, we adjusted the systems' dimensions to sample the configurational space of the unstrained sheets in the canonical ensemble. 
In an analogous manner, we adapted the box dimensions to achieve relative strains between $-2.0$~$\%$ and $+2.0$~$\%$ to study the shape of mechanically induced ripples. 
The sampling time for all considered strains was roughly 0.5 ns.\\

\section{Results and Discussion}
\subsection*{\hl{Evaluation of the hBN-GAP}}
We validate the hBN-GAP and benchmark its predictive capabilities against its DFT reference, comparing it to other frequently used force fields, specifically the Tersoff potential \cite{Tersoff1986, Tersoff1988} parametrised by Sevik et al. \cite{Sevik2011}, the extended Tersoff potential (ExTeP) employing a modified bond-order parameter \cite{Los2017}, and the ReaxFF \cite{VanDuin2001,Chenoweth2008} parametrised by Weismiller et al. \cite{Weismiller2010}. In addition to the standard versions, we also evaluated the predictions made by both Tersoff and ExTeP when combined with the recently developed interlayer potential (ILP) \cite{Leven2014,Leven2016,Maaravi2017}. This coupling enables these models to account for the intermolecular interactions between hBN sheets. These are typically neglected by the independent bond-order potentials due to a cut-off of 2.0 $\mathrm{\AA}$ which is below the equilibrium separation between layers in bulk hBN of about 3.3 ${\mathrm{\AA}}$. The list of force field models tested is by no means exhaustive and is merely meant to serve as baseline for our newly developed model in the context of previous work. \hl{This is also the case for the physical properties we evaluate the performance of the developed model on.} We also note that all DFT comparisons employ the very same functional, \hl{PBE+D3}, and electronic structure setup as applied in generation of training data. While this is essential to appropriately evaluate the quality of the constructed hBN-GAP, it might distort the benchmarking by implying that established force fields are in error when deviating from DFT results based on our chosen functional. To go some way to address this bias, we compared the performance of different functionals -- local density approximation (LDA) and other generalised gradient approximations (GGA) -- for basic crystalline properties of hBN (see supplementary information for details).\\

\begin{table*}[ht!]
    \caption{Comparison of lattice parameters and interlayer distances for the different hBN phases between DFT (\hl{PBE+D3}), the hBN-GAP, and other force fields. The lattice parameters are given in absolute values while the relative error with respect to DFT is given in percentage in brackets. For bulk hBN the out-of-plane lattice parameter c corresponds to the length of the respective lattice vector while for an isolated bilayer $\mathrm{\tilde{\mathbf{c}}}$ is defined as the interlayer distance. Due to the short cut-off of Tersoff and ExTeP no values (\---) are given for the out-of-plane lengths of bilayer and bulk hBN. }
    \label{tab:geometrical_properties}
    \begin{adjustbox}{center} 
        \begin{tabular}{lccccccc}
             \toprule
             & \multicolumn{7}{c}{Lattice Parameter [$\mathrm{\AA}$] (\% Error)} \\ 
             \cmidrule(r){2-8}
             & {DFT} & {GAP} & {Tersoff} &{Tersoff + ILP} & {ExTeP} & {ExTeP + ILP}& {ReaxFF}  \\
             \midrule
            Bulk hBN (a)       & 2.51 & 2.51 (0.0) & 2.50 (0.4)  & 2.50 (0.4) &  2.50 (0.4) & 2.50 (0.4) & 2.55 (1.6) \\ 
            Bulk hBN (c)       & \hl{6.61} & \hl{6.60 (0.2)} & \---  & 6.46 \hl{(2.3)} &   \--- & 6.45 \hl{(2.4)} &5.88 \hl{(11.0)} \\ 
            Bilayer (a)        & 2.51 & 2.51 (0.0)  &2.50 (0.4) & 2.50 (0.4) &  2.50 (0.4)& 2.50 (0.4)   & 2.55 (1.6)    \\
            Bilayer ($\tilde{\mathrm{c}}$)        & \hl{3.36}& \hl{3.36 (0.0)} & \---  & 3.26 \hl{(3.0)} & \--- & 3.26 \hl{(3.0)} & 2.99 \hl{(11.0)}   \\ 
            Monolayer (a)      &  2.51  & 2.51 (0.0) & 2.50 (0.4) & 2.50 (0.4) &  2.50 (0.4) & 2.50 (0.4) & 2.55 (1.6)   \\ 
            Nanotube (10,0) & 4.34 & \hl{4.32 (0.5)}   &  4.32 (0.5)  &4.32 (0.5) & 4.32 (0.5) & 4.32 (0.5) &   4.34 (0.0)  \\ 
            Nanotube (10,10) &  2.51 & 2.51 (0.0) & 2.50 (0.4) & 2.50 (0.4)  & 2.51 (0.0) &2.50 (0.4)  & 2.55 (1.6) \\
            
            \bottomrule
        \end{tabular}
    \end{adjustbox}
    \label{tab:crystalline_properties}
\end{table*}

The most intuitive measure to evaluate the performance of a ML potential is through force and energy errors on a basis of a validation set of configurations. This analysis has been carried out for most existing GAP models describing a great variety of materials \cite{Bartok2010,Deringer2017,Rowe2018}. Force correlation plots as well as a comparison of computational costs for DFT, GAP, and the established models can be found in the supplementary information. Here, we will only report the root mean square errors (RMSE) with respect to DFT for the different models. For a validation set of 1450 randomly picked structures\hl{, which were not included in the training,} the hBN-GAP achieves a RMSE of \hl{0.09}~$\mathrm{eV}/\mathrm{\AA}$. The established potentials are one order of magnitude less accurate with Tersoff, ExTeP, and ReaxFF yielding RMSEs of 1.09, \hl{1.19}, and \hl{6.17}~$\mathrm{eV}/\mathrm{\AA}$, respectively. The ILP extended bond-order potentials perform surprisingly slightly worse with \hl{1.11} and \hl{1.22}~$\mathrm{eV}/\mathrm{\AA}$ for Tersoff+ILP and ExTeP+ILP, respectively. While a correct reproduction of the atomic forces is a crucial requirement for any accurate interatomic potential, it is, however, not a sufficient criterion to secure an accurate description of the system of interest. In the course of this work, we observed that previously developed versions of our potential showed similar force errors but diverged significantly when comparing material properties. \hl{Moreover, high force errors do not lead to a bad description of macroscopic properties \textit{per se}.} Therefore, the \hl{evaluation} presented will be focused on how the GAP model performs for energetic, geometrical, mechanical, and vibrational properties of different hBN phases. \\

The reproduction of lattice parameters for different phases is an essential requirement an accurate potential must fulfil. A comprehensive comparison of lattice parameters predicted by the hBN-GAP and other force fields as well as the relative deviation from DFT can be found in table \ref{tab:geometrical_properties}. The DFT results agree well with experimental results \cite{Solozhenko1995} for the \hl{monolayer and }bulk phase. Overall, our model performs very well with respect to its DFT reference \hl{for all lattice lengths and structures with an average error of 0.10~\%.}  \\

The established force fields considered also achieve a very good performance for the geometric properties albeit being slightly less accurate than our hBN-GAP. Tersoff and ExTeP yield average errors of 0.42~\% and 0.34~\%, respectively, although their actual score is somewhat biased as no out-of-plane lengths were considered. Coupling both potentials tuned with ILP, however, allows for the description of multilayered hBN and they both yield an average error of \hl{1.57~\% and 1.59~\%, respectively}. ReaxFF achieves a sufficient accuracy for the in-plane and nanotubes lattice parameters but it fails dramatically to capture the out-of-plane lengths resulting in an overall error of \hl{ 4.06~\%}.\\

\begin{figure}[ht!]
    \centering
    \begin{tikzpicture}
    \node (fig) at (0,0) {\includegraphics[width=\linewidth]{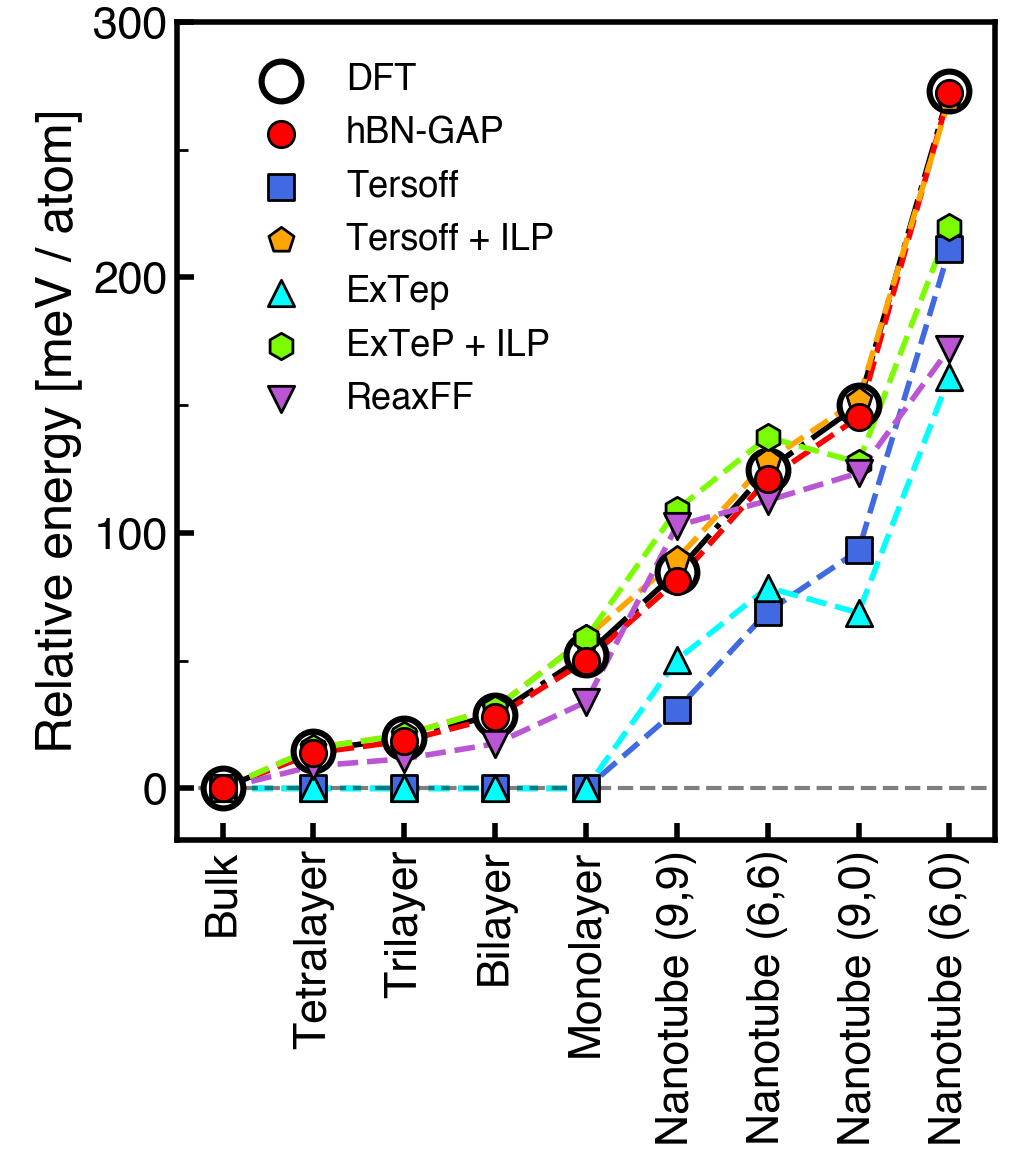}};
     \end{tikzpicture}
    \caption {Formation energies of the hBN configurations computed with DFT (\hl{PBE+D3}), the hBN-GAP, and other established force fields. Both bond-order potentials Tersoff and ExTeP are shown with and without the correction for multiple layers contributed by the ILP. All energies are plotted relative to bulk hBN. The lines are a guide to the eye.} 
    \label{fig: relative energies}
\end{figure}

In addition to the structural characteristics, we now consider the energetics of the relevant phases as shown in figure~\ref{fig: relative energies}. To this end, the computed formation energies of the optimised structures are given per atom and relative to bulk hBN which is the thermodynamically most stable phase. In the case of layered hBN, the observations from the lattice parameters for Tersoff and ExTEP are confirmed. Due to the lack of non-bonded contributions to the energy these potentials cannot distinguish between different numbers of sheets yielding an RMSE of more than \hl{54 and 67~meV/atom, respectively}. Compensating for this with the ILP, the tuned Tersoff and ExTeP potentials achieve \hl{very good agreement with DFT with an improved RMSE of 4 and 26~meV/atom}, respectively. While the performance of ReaxFF is mediocre with an error of 45~meV/atom, our hBN-GAP achieves also \hl{very good agreement with DFT with a RMSE of 3~meV/atom}. This high accuracy is persistent throughout the tested configurations including the energetically less favourable nanotubes irrespective of diameter or chirality. To avoid confusion it is worth mentioning that the total energy of the nanotubes predicted by Tersoff and ExTeP is identical to those predicted by their versions enhanced by ILP. The energy shift in the plot is caused by the change of the energy for the reference bulk hBN.\\

Part of the attention garnered by hBN is due to its excellent mechanical properties. Moreover, a system's response to mechanical compression and tension is an essential property for 2D materials as it affects the shape of ripples in atomically thin sheets \cite{Cerda2003,Fasolino2007}. In order to ensure an accurate description of the strain-deformation relationship we computed the relevant elastic constants of bulk hBN with DFT, GAP, and existing potentials. An overview of these results and a comparison to experimental measurements \cite{Bosak2006} can be found in table \ref{tab: elastic constants}. In order to determine the elastic constants, finite distortions were made to the optimised cell and the ions were allowed to relax. The respective constants were then calculated from the stress-strain relationship whereby different magnitudes of strain were applied to guarantee a linear behaviour. For ReaxFF this linear relationship could not be achieved. Instead physically unrealistic and negative elastic constants were obtained. This indicates that the parametrisation of the ReaxFF used \cite{Weismiller2010} does not yield a satisfactory description of the equilibrium shape of bulk hBN, namely AA$^\prime$ stacking in a hexagonal unit cell. This issue could not be solved by changing to a different parameter set \cite{Liu2017} for ReaxFF. The results for ReaxFF are, thus, not included in table \ref{tab: elastic constants}.\\

\begin{table*}[t]
    \caption{Mechanical properties of bulk hBN calculated with DFT (\hl{PBE+D3}), the hBN-GAP, and other force fields in comparison to experimental measurements \cite{Bosak2006}. For the experiments the values in parenthesis correspond to the experimental uncertainty. Due to the short cut-off of Tersoff and ExTeP no values (\---) are given for the out-of-plane elastic constants C$_{13}$, C$_{33}$, C$_{44}$, and the bulk modulus.}
    \begin{adjustbox}{center} 
        \begin{tabular}{lccccccc}
             \toprule
             & \multicolumn{7}{c}{Elastic constants [$\mathrm{GPa}$]} \\ 
             \cmidrule(r){2-8}
             & {DFT} & {GAP} & {Tersoff} & {Tersoff + ILP} & {ExTeP} & {ExTeP + ILP}& {Experiment \cite{Bosak2006}} \\
             \midrule
             C$_{11}$       & \hl{881.8} & \hl{853.0} & 969.9  & 858.0  & 849.1 & 861.9  &    811.0 (12.0) \\ 
            C$_{12}$        & \hl{195.6} & \hl{216.8} & 303.0  & 269.8  &152.4 & 156.3  &  169.0 (24.0) \\ 
           C$_{13}$     & 0.0 & \hl{12.9}  &\--- & 2.8 &\--- & 2.8&  0.0 (3.0)\\
            C$_{33}$     & \hl{28.9} & \hl{36.1}  &\--- & 37.9  &\---& 37.7  & 27.0 (5.0)     \\ 
            C$_{44}$    &  \hl{3.3} & \hl{0.3} &  \--- & 6.5  & \--- &  6.5  &   7.7 (5.0)\\ 
            B (eq) & \hl{27.4} & \hl{35.4}  & \--- & 35.9 & \---&  35.4  & \multirow{2}{*}{25.6 (8.0)} \\ 
            B (MEoS) & \hl{31.2} & \hl{34.7} &\--- &39.8 & \--- &  39.7 &  {}\\ 
                \bottomrule
        \end{tabular}
    \end{adjustbox}
    \label{tab: elastic constants}
\end{table*}

Our DFT calculations match the measurements for most constants within the experimental uncertainty. Both approaches predict that there is no coupling between in-plane stress (strain) and out-of-plane strain (stress), i.e. $\mathrm{C}_{13} = 0$. Our hBN-GAP achieves a very good agreement with its reference DFT, particularly for the in-plane constants $\mathrm{C}_{11}$, $\mathrm{C}_{12}$ and the out-of-plane coupling $\mathrm{C}_{33}$ which also corresponds to a high accuracy of the interlayer binding curve \cite{Graziano2012}. \hl{The prediction for $\mathrm{C}_{13}$ and $\mathrm{C}_{44}$ is less accurate while the actual behaviour, i.e. a weak coupling between stress and strain in the respective directions, is still captured qualitatively. It is worth noting, that very tight target deviations for the virials $\sigma_\mathrm{virial}$ are required to obtain very high accuracy for these quantities, however, using such target deviations can lead to overfitting.} Beyond the individual couplings between different directions, a correct description of the overall compressibility of the material, i.e. the bulk modulus $\mathrm{B}$, is desirable. Here, we calculate $\mathrm{B}$ via a relationship \cite{Ohba2001} dependent on the elastic constants, 
\begin{equation}
  \mathrm{B}=\frac{\mathrm{C}_{33}(\mathrm{C}_{11}+\mathrm{C}_{12})-2(\mathrm{C}_{13})^2}{\mathrm{C}_{11}+\mathrm{C}_{12}+2\mathrm{C}_{33}-4\mathrm{C}_{12}} \mbox{ , }
\end{equation} and by fitting the Murnaghan equation of state (MEoS) \cite{Murnaghan1944} to the energy-volume curve. For the latter approach, the anisotropic character of hBN requires a non-isotropic compression to avoid erroneous results. As shown in table \ref{tab: elastic constants}, both methods yield similar results and the hBN-GAP agrees well with DFT.\\

Evaluating the performance of established potentials, both Tersoff and ExTeP suffer from their unsatisfactorily short cut-off predicting infinite compressibility for any reasonable geometry. Therefore, no values are included for elastic constants dependent on out-of-plane strain. By applying the ILP correction both tuned potentials perform very well compared to DFT and experiments. While they are in average sligthly less accurate for the in-plane constants they show an even better agreement for $\mathrm{C}_{13}$ and $\mathrm{C}_{44}$ than the hBN-GAP. Despite this good performance, it should be noted that the high accuracy seems to be mainly based on the ILP correction. Therefore, doubts must be cast on the predictive capability of these potentials for isolated hBN sheets and nanotubes as the ILP has no contribution to the potential energy in these cases.\\

For low-dimensional materials, it has been shown that phonon modes affect adsorbate (notably water) diffusion through carbon nanotubes \cite{Ma2015} and across graphene \cite{Ma2016}. Other relevant thermodynamic properties, such as heat capacity, thermal conductivity, and thermal expansion coefficients are also closely linked to phonons. Therefore, an accurate description of the dynamics of the lattice is an essential requirement for the hBN-GAP.\\

\begin{figure}[ht]
    \begin{tikzpicture}[align=center]
    \node (fig) at (0,0)  {\includegraphics[width=0.99\linewidth]{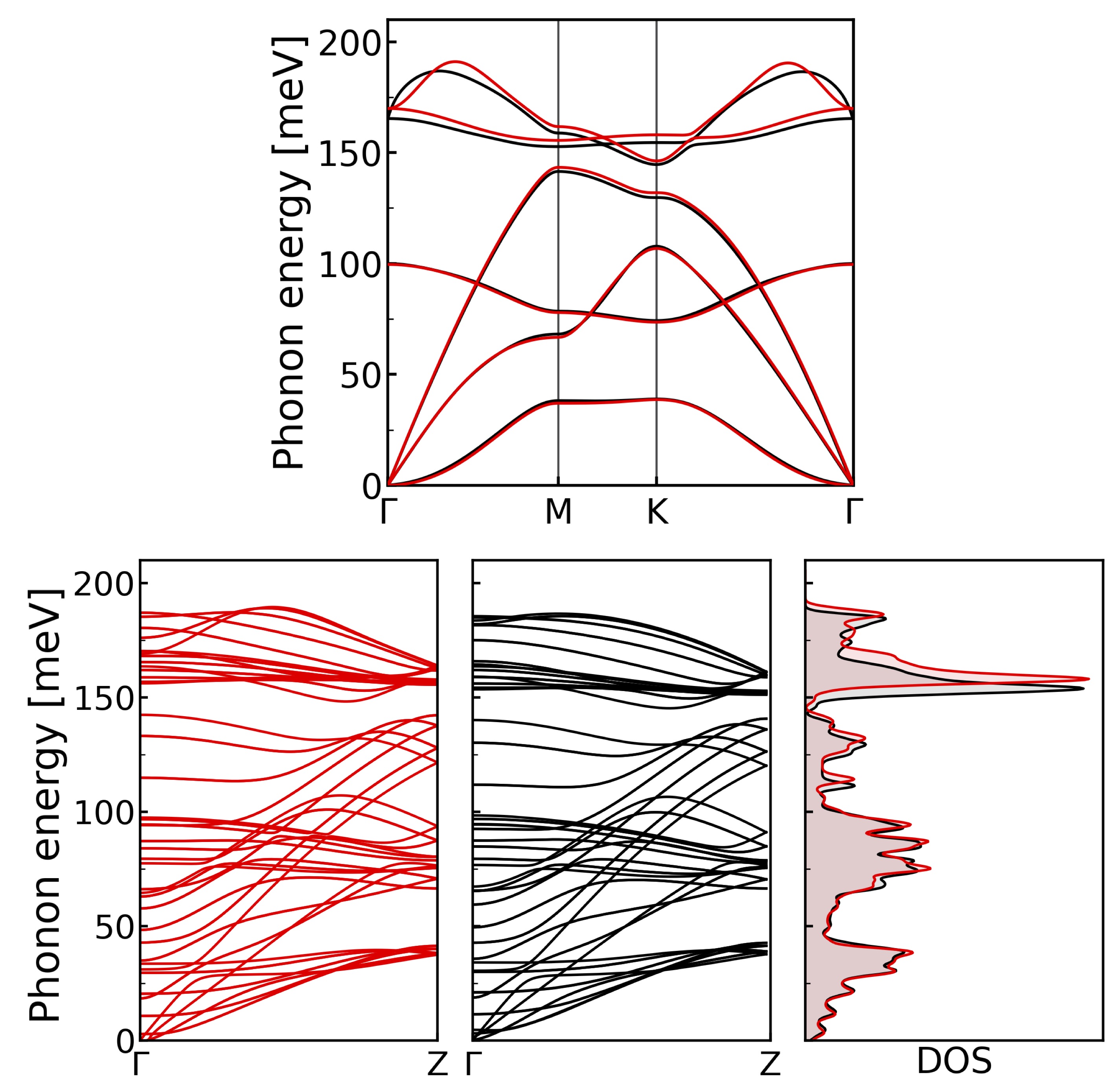}};

    \node at (-2.4,3.8) {\textbf{A} }; 
      \node at (-3.9,0.05) {\textbf{B} };

    \end{tikzpicture}
    
    \caption [test]{Phonon dispersion curves calculated for different phases of hBN. The black data corresponds to the DFT reference\footnotemark (\hl{PBE})  while the red data represents our hBN-GAP. (\textbf{A}) Phonon dispersion for an isolated hBN sheet. (\textbf{B}) \hl{Phonon dispersion for a selected nanotube with chirality (6,6) as well as the density of states (DOS) predicted with GAP and DFT.} Due to the number of bands, the phonon dispersion for other force fields are included in the supplementary information.}
    \label{fig: phonons}
\end{figure}

\footnotetext{\hl{As the D3 dispersion correction is not implemented in the DFPT framework of the QE software package, the phonon calculations of the hBN monolayer and nanotube are done using the PBE functional without dispersion correction.}}

Before discussing the performance of the hBN-GAP, it is important to emphasise some essential aspects in the calculation of the phonon dispersion curves for hBN. \hl{Due to the material's polar nature, a dipole is induced by the longitudinal optical (LO) phonon modes which creates a macroscopic electric field affecting the dispersion behaviour of the optical phonons in the long wavelength limit, i.e. $q \rightarrow 0$. This complicates the determination of phonon dispersion curves within the so-called finite displacement method requiring an infinitely large supercell {\cite{Alfe2009}} or the addition of a non-analytical correction to the dynamic matrix based on the Born effective charges {\cite{Giannozzi1991}} as shown by Gonze et al. {\cite{Gonze1994,Gonze1997}}.} Alternatively, the analytical part of the dynamical matrix is also readily calculated by the linear response of the electron density providing the entire dispersion curve based on the unit cell only \cite{Baroni2001}.\\

Here, as first-principle reference we conduct density functional perturbation theory (DFPT) calculations \hl{because an appropriate treatment of the Coulombic interactions for 2D materials} has recently been implemented \cite{Sohier2017} in the DFPT framework in the QUANTUM ESPRESSO (QE) \cite{Giannozzi2009,Giannozzi2017} software package. Equivalent setups for the DFPT and DFT calculations conducted with QE and VASP, respectively, were used. We ensured a high agreement between both codes throughout different properties and a more detailed comparison can be found in the supplementary information. For GAP and other potentials the phonon dispersion curves are calculated using the finite displacement method provided by the Phonopy package \cite{Togo2015}. For potentials treating charges explicitly, contributions due to the non-analytical correction according to Gonze et al. \cite{Gonze1994,Gonze1997} can be added. This, however, does not apply for the bond-order potentials and the hBN-GAP. \\

In this work, we compare the phonon dispersion of a \hl{hBN single layer and a hBN nanotube with chirality (6,6).} \hl{Very good} agreement was obtained between our \hl{hBN-GAP} and inelastic X-ray scattering (IXS) measurements \cite{Geick1966,Nemanich1981,Reich2005,Serrano2007} for bulk hBN which is reported in the supplementary information. The dispersion curves computed with DFPT and hBN-GAP are shown in figure~\ref{fig: phonons} and the performance of other force fields can be found in the supplementary information. As shown in figure \ref{fig: phonons}A the GAP achieves a very good agreement with its DFT reference, particularly for the acoustic modes. The predicted frequencies of the optical branches at the $\Gamma$-point are correct to $<5$~meV. GAP is not able to reproduce the finite slope of the LO dispersion close to the zone center. The generated electric field causes a non-zero slope which can be captured by adding the non-analytical correction to the dynamic matrix as described above \cite{Sohier2017}. As charges and related electrostatics are treated implicitly, the current version of the hBN-GAP cannot reproduce this characteristic behaviour. \hl{It is worth noting, however, that explicit charges can be incorporated in the GAP framework which will be part of future extensions of the hBN-GAP.}\\

The good performance of the hBN-GAP for the vibrations in \hl{monolayer} hBN becomes even more apparent when comparing to existing force fields shown in the supplementary information. Both the Tersoff and ExTeP potentials provide a good though less accurate description of the acoustic modes yielding deviations of more than 20 meV. Further, the phonon energy of the optical modes, particularly LO and TO, is overestimated by up to 40 meV for both potentials. \\

To understand the performance of the hBN-GAP for a (6,6) nanotube, we focus on the overall phonon density of states (DOS) rather than individual bands due to the large number of atoms in the unit cell. As depicted in figure \ref{fig: phonons}\hl{B}, the hBN-GAP reproduces the DFT results very well showing only small deviations of typically \hl{$<5$ meV}. The DOS predicted by the hBN-GAP is almost congruent with the one obtained from DFT calculations \hl{except for a small offset for high energy modes. However, these deviations are in the same order of magnitude as the difference between predictions made by different DFT functionals {\cite{Wirtz2003}}}. For the established force fields we observe a similar trend to layered hBN. While the low energy modes are described sufficiently accurately yielding deviations of about 7 meV, the high energy modes shifted by more than 30 meV for both Tersoff and ExTeP potential (see supplementary information).

\subsection*{Application to rippling in hBN and graphene}



After the careful validation in the previous section, we focus on the application of the hBN-GAP to investigate rippling in single layer hBN in comparison to graphene.
Although their name may suggest otherwise, these 2D materials are not perfectly flat at finite temperature but exhibit local height fluctuations \cite{Fasolino2007,Zakharchenko2010,Los2009,Slotman2013,Katsnelson2013,Bao2009,Xu2014}.
\hl{These thermally activated ripples are an intrinsic feature of 2D crystals which are stabilised by an anharmonic coupling between bending and stretching modes} \cite{Nelson1987,LeDoussal1992,Nelson2004,Cerda2003,Meyer2007}.
Moreover, the corrugation of the surface can strongly affect the material's properties  \cite{Guinea2008,Guinea2008a,VazquezDeParga2008} indicating a close link between atomic and electronic structure. 
The behaviour of the 2D crystal is, thus, directly affected by the average amplitude of the height fluctuations, $\langle h\rangle$, which, according to the theory of flexible membranes \cite{Nelson1987,LeDoussal1992,Nelson2004}, scales with the system size, $L$, as $\langle h\rangle \propto L^{1-\eta/2}$, with $\eta$ being the anomalous rigidity exponent. While the results of atomistic simulations on graphene \cite{Los2009} and a general nonperturbative renormalisation-group approach \cite{Kownacki2009} suggest the transferability of $\eta=0.85$, yet,\hl{ a smaller value of $\eta \approx 0.66$ was obtained for hBN {\cite{Los2017}} based on molecular dynamics simulations using the ExTeP model. This deviation, however, may stem from simulation settings, particularly, too small length and time scales. While system sizes required to confirm this assumption exceed 50000 atoms and are, thus, beyond the scope of this work, valuable insight on the predictive capability of the hBN-GAP can be gained based on smaller systems comprising 7200 atoms.\\}

Here, we first utilise the hBN-GAP to determine $\eta$ based on MD simulations following the procedure described by Los et al. \cite{Los2009}. For the sake of comparability, we also compute $\eta$ for graphene using the same methodology whereby the interatomic interactions are described by the graphene-GAP potential \cite{Rowe2018}.
However, as the actual shape and amplitude of the ripples can be strongly altered by externally introduced strain \cite{Bao2009,Xu2014,Ma2016}, we subsequently analyse the impact of tension and compression on the mean amplitude and phase behaviour of both materials.\\

The scaling exponent $\eta$ is computed based on the comparison between the results of our atomistic simulations and the prediction of the normal-normal correlation $G(q)$ function provided by the theory of flexible membranes\cite{Nelson1987,LeDoussal1992,Nelson2004}.
A general expression for $G(q)$ is given by the Dyson equation
\begin{equation}
  G^{-1}(q)=  G_0^{-1}(q) + \Sigma(q) \mbox{ , }
  \label{eq:dyson}
\end{equation}
where $G_0(q)$ represents the harmonic approximation,
\begin{equation}
  G_0(q)=  \frac{TN}{\kappa S_0 q^2} \mbox{ , }
  \label{eq:harmonic}
\end{equation}
and $\Sigma(q)$ is the self energy accounting for the anharmonic coupling at small wavevectors,
\begin{equation}
  \Sigma(q)=  \frac{AS_0}{N}q^2\left(\frac{q_0}{q}\right)^\eta \mbox{ , }
  \label{eq:self_energy}
\end{equation}
with $q$ as length of the 2D wavevector $\mathbf{q}$, $T$ as temperature in energy units, $N$ as number of atoms, $\kappa$ as bending rigidity, $S_0$ as area per atom, A as an unknown prefactor, $q_0=2 \pi \sqrt{B/\kappa}$, B as 2D bulk modulus, and $\eta$ as the desired scaling exponent. In the limit of slowly varying height fluctuations ($\left|\nabla h\right|^2 \ll 1$), $G(q)$ as defined in equation \ref{eq:dyson} is identical to $q^2 \langle \left| h(q) \right|^2 \rangle$, where  $h(q)$ represents the Fourier transform of the atomic out-of-plane displacement, $h(\mathbf{x})$, which is directly obtained from a MD trajectory \cite{Fasolino2007,Los2009,Zakharchenko2010}.
In practice, however, $h(\mathbf{x})$ is smoothed by averaging over the nearest neighbours before numerically calculating the Fourier components \cite{Zakharchenko2010}.
Further, the wavevectors are restricted by the system size to a minimum length of $2 \pi / L$ corresponding to $\approx 0.048$ $\mathrm{\AA}^{-1}$ for our system, which has been shown to be sufficiently small to access the anharmonic region \cite{Los2009}.\\
\begin{figure}[ht!]
\centering
    \begin{tikzpicture}
    \node (fig) at (0,0) {\includegraphics[width=\linewidth]{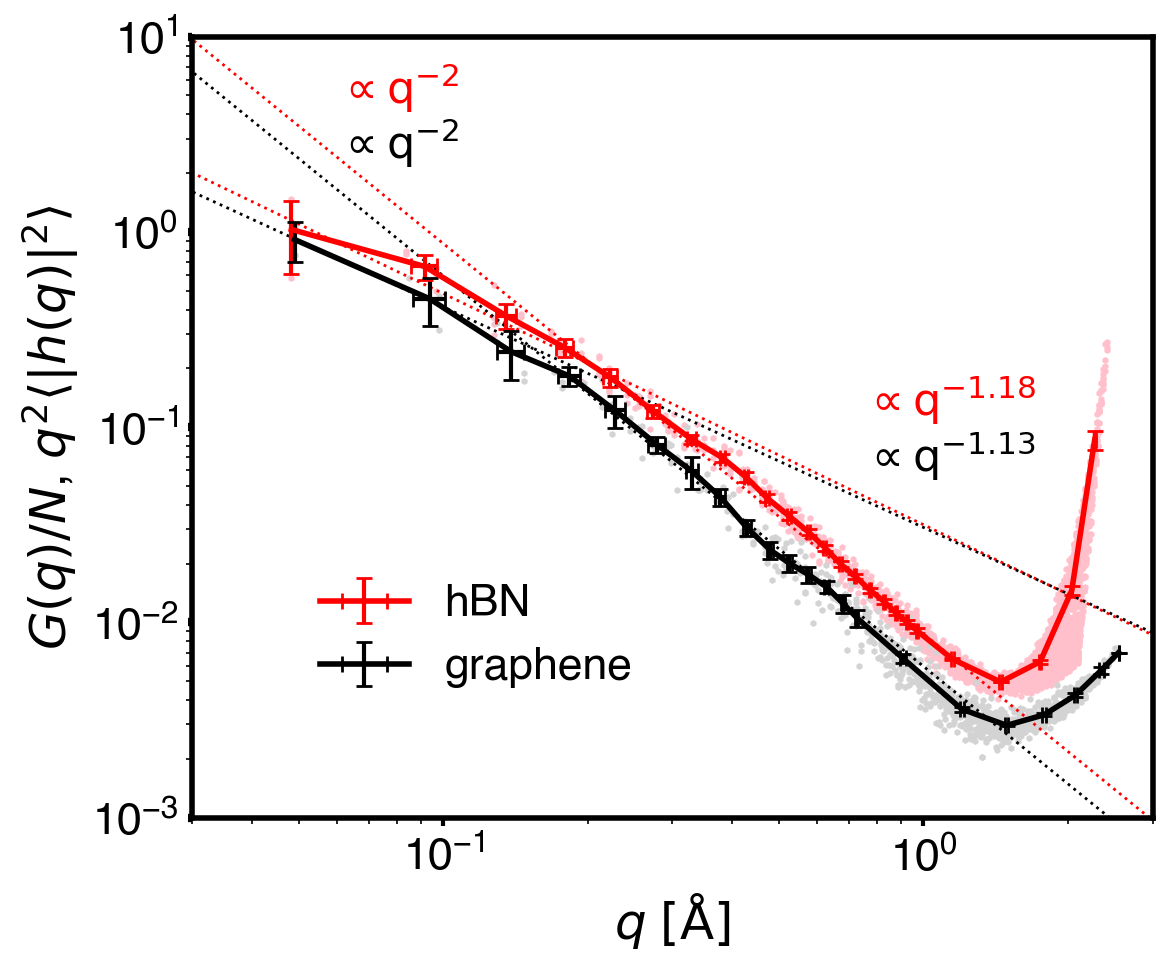}};

     \end{tikzpicture}
 
    \caption {Normal-normal correlation function $G(q)/N$ calculated for graphene and hBN based on MD simulations using the hBN-GAP and the graphene-GAP, respectively. The pale markers correspond to the numerical results directly obtained from our simulations. This data was then smoothed and averaged within small intervals of wavevectors which is represented by the continuous line including error bars. The dotted lines imply the asymptotic behaviour with power-law $q^{-2}$ for the harmonic approximation for large $q$ and $q^{-(2-\eta)}$ for long wavelengths where the anharmonic contributions dominate. For graphene and hBN values of $\eta=0.87$ and \hl{$\eta=0.82$}, respectively, are obtained} 
    \label{fig: corr_func}
\end{figure}

Figure \ref{fig: corr_func} depicts $G(q)/N$ calculated from the MD trajectories for unstrained hBN and graphene. While for large wavevectors the continuum theory breaks down, $G(q)$ is accurately reproduced by the harmonic approximation $G_0(q)$ for wavevectors in range $q^* \lesssim q  \lesssim 1$~$\mathrm{\AA}^{-1}$ where $q^*$ is given by the Ginzburg criterion \cite{Nelson2004}
\begin{equation}
    q^* = \sqrt{\frac{3TB}{8\pi\kappa^2}} \mbox{ , }
    \label{eq:q_star}
\end{equation}
quantifying the long-wavelength limit where anharmonic contributions become dominant.
We extracted the bending rigidity $\kappa$ at 300~K for both materials by comparing equation \ref{eq:harmonic} to the simulation results for wavevectors $0.3 < q < 1$~$\mathrm{\AA}^{-1}$ yielding \hl{$\kappa = 1.09$}~eV and $\kappa = 1.65$~eV for hBN and graphene, respectively. 
As these values seem very high in comparison to previous work using classical potentials \cite{Fasolino2007,Los2009,Slotman2013}, we computed $\kappa$ at 0~K based on the out-of-plane acoustic phonon branch (ZA) \cite{Karssemeijer2011} for both GAP models. 
The results of $\kappa = 0.90$ eV and $\kappa=1.58$ eV for hBN and graphene, respectively, agree very well with previous DFT calculations \cite{Wu2013,Wei2013}.\\

We now focus on the long-wavelength simulation data in the range $q\leq q^*$ which we compare to the anharmonic self energy $\Sigma(q)$ given by equation \ref{eq:self_energy}. 
Due to the rather weak temperature dependence of the 2D bulk modulus $B$ \cite{Zakharchenko2009}, we used the predicted values of the GAP models for 0~K  which are \hl{$B=11.38$~eV/$\mathrm{\AA}^2$} and $B=13.59$~eV/$\mathrm{\AA}^2$ for hBN and graphene, respectively.
Further, to be comparable with previous work \cite{Fasolino2007,Los2009} we fixed the prefactor $A=1$. 
This way, a best fit to the simulation data yields scaling exponents of \hl{$\eta=0.82$} for hBN and $\eta=0.87$ for graphene which agree very well with established findings \cite{Los2009,Kownacki2009}.

hl{As pointed out above,} the small deviations between the materials might be caused by the limited availability of fitting data in the long-wavelength range \hl{requiring significantly larger system sizes.}
Our results, therefore, confirm the observation made by Los et al. \cite{Los2009} that the scaling exponent is independent of the bending rigidity $\kappa$.
Conversely, the distinct materials' properties strongly affect the actual height of the ripples where small values of $\kappa$ result in larger height fluctuations as shown by the shift of hBN in figure \ref{fig: corr_func}.
Also, according to equation \ref{eq:q_star} the crossover between the harmonic and anharmonic regime is switched to lower $q$ for lower values of $B/\kappa^2$ which is also confirmed by our results.\\

So far, we have shown that the scaling behaviour of ripples \hl{is predicted to be almost} identical in hBN and graphene while the amplitude of the out-of-plane deviations for a given system size will be slightly larger in hBN than in graphene.
However, these findings apply in the limit of no external strain. 
To go beyond this, we analyse the average height of these mechanically induced ripples based on  MD trajectories of hBN and graphene exposed to different intensities of strain.
To this end, we determined the atom which shows the highest deviation from the centre mass perpendicular to the flat sheet.
This measure is then averaged over the entire trajectory and the respective error based on block averages was computed to quantify the uncertainty. 
We also evaluated the influence of the box geometry on the shape and amplitude of the ripples in compressed systems. 
While the shape changes significantly from a hexagonal to an orthorombic simulation box the magnitude of the height fluctuations is almost unaffected (see supplementary material).\\

\begin{figure}[h!]
   \centering
    \begin{tikzpicture}
    \node (fig) at (0,0) {\includegraphics[width=\linewidth]{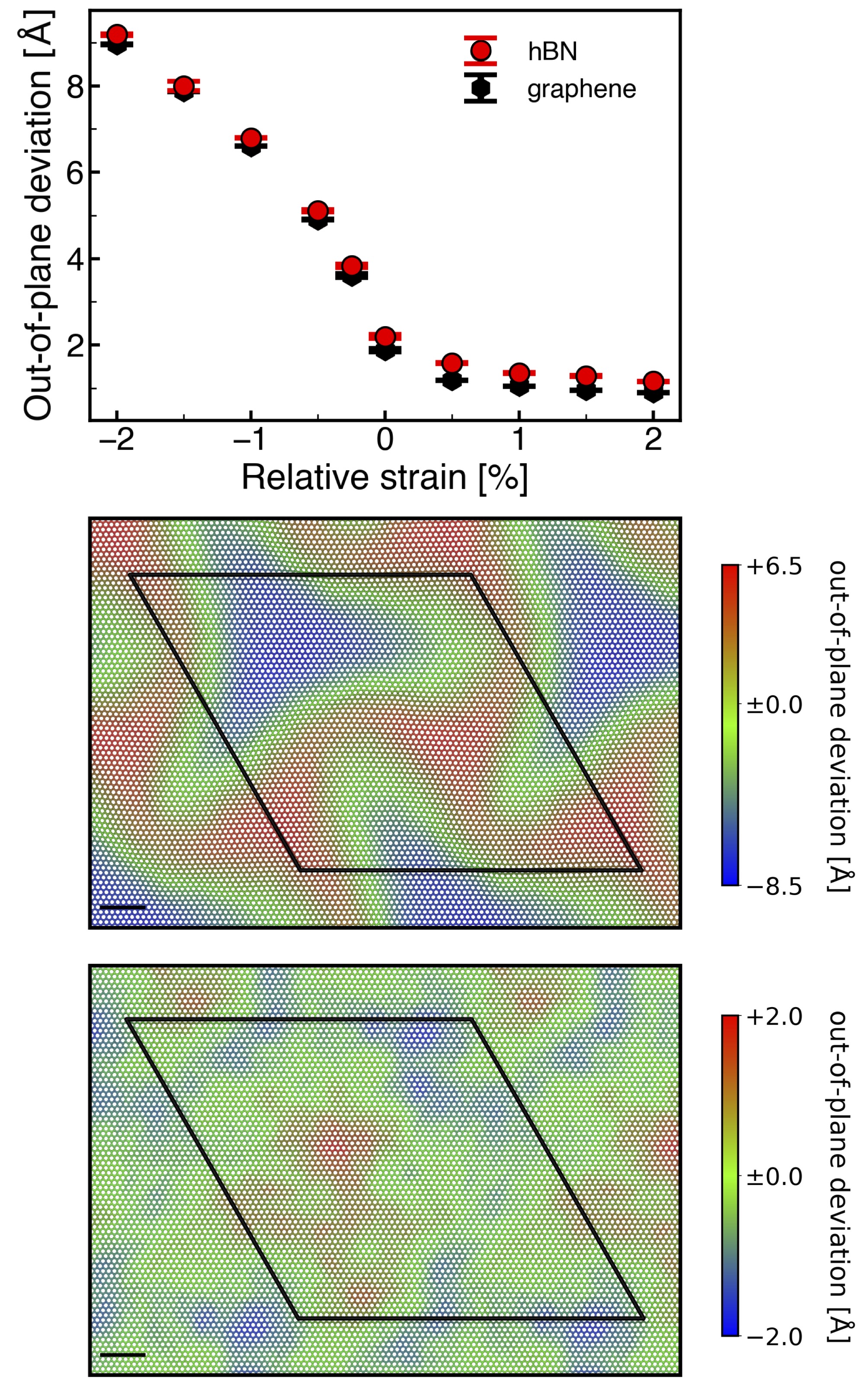}};
    
    \node (A) at (-3.9,7.2) {\large{\textbf{A}}};
    \node (B) at (-3.9,2.0) {\large{\textbf{B}}};
    \node (C) at (-3.9,-2.5) {\large{\textbf{C}}};

     \end{tikzpicture}
    \caption {(\textbf{A}) Averaged out-of-plane deviation as function of applied strain for hBN (red) and graphene (black) predicted by the hBN-GAP and graphene GAP, respectively.(\textbf{B-C}) Snapshot of the MD simulations for an unstrained and hBN monolayer and compression of $-2.0\%$. The atoms are coloured according to their displacement from the center of mass in z-direction. The simulation box is indicated in black and the scale bar in the bottom left represents 2~nm.}
    \label{fig:rippling}
\end{figure}
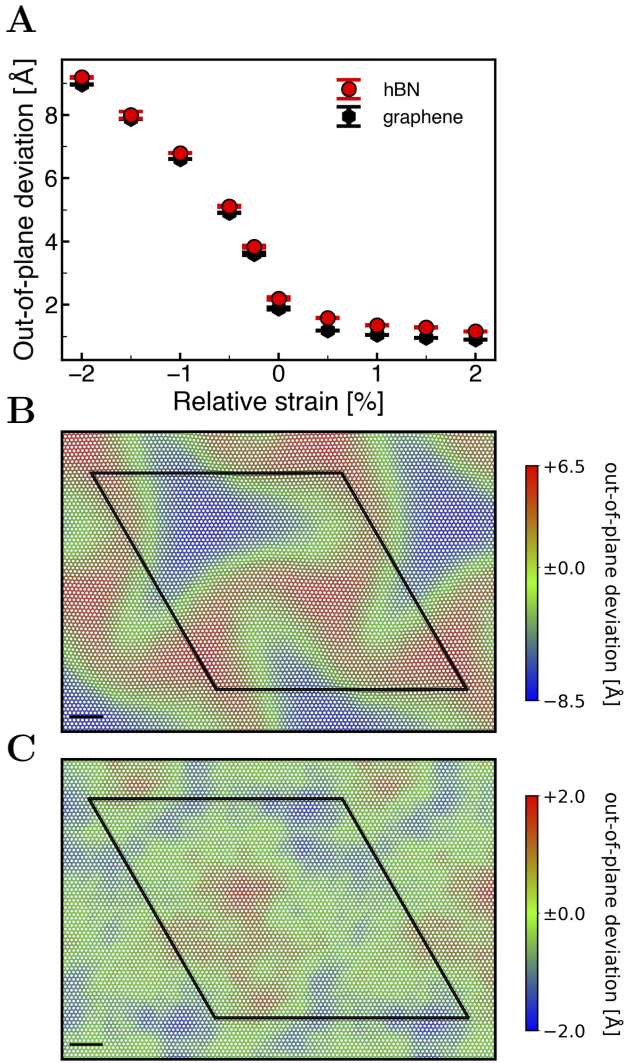

In analogy to our findings for the unstrained systems, ripples in hBN and graphene show a very similar reaction to strain as shown in figure \ref{fig:rippling}A. 
The measured amplitudes agree with previous simulation work based on established force fields for both graphene  \cite{Ma2016} and hBN \cite{Thomas2017}.
Throughout the entire strain range, hBN shows a larger average rippling height than graphene due to its lower bending rigidity $\kappa$.
Large negative strains lead to a compression of the 2D sheets as the system stabilises by deforming in the third dimension while uniaxial tension causes the amplitude of the out-of-plane deformations to shrink.
We observed a phase transition from short-lived fluctuations (figure \ref{fig:rippling}B) to spatially coherent soliton-like ripples (figure \ref{fig:rippling}C) which takes place between a strain of $-0.25\%$ and $-0.5\%$ for both materials. 
This is in line with simulations performed on isolated graphene sheets \cite{Ma2016}.
Interestingly, for multilayered graphene previous simulations \cite{DeLima2015} have led to the suggestion that a much higher strain of $-2.8\%$ is required to induce spatially coherent ripples.
Exploring this issue for hBN and the impact of vdW forces between multiple layers on the phase transition of ripples is a fascinating field and will be subject of future work. \\

Finally, it is worth mentioning that these studies on ripples show that the hBN-GAP proves to be transferable to system sizes outside both the training set and the accessible scale of DFT.
Indeed, the predicted phase transition from figure \ref{fig:rippling}B to a rippling pattern as in figure \ref{fig:rippling}C is exclusively based on an interpolation between the local environments in the training set.

\section{Conclusion}
In this work, we introduced a machine learning-based atomistic model to treat different phases of hBN ranging from isolated sheets to the bulk phase and nanotubes of any chirality. 
Our model was constructed using the GAP methodology and was trained on energies, forces, and virial stresses obtained from tightly converged vdW-inclusive DFT calculations.
We benchmarked the hBN-GAP against its DFT reference and other established potentials and demonstrated its capabilities through a variety of tests including the elastic constants and the phonon dispersion spectra for nanotubes and layered hBN. 
Based on this \hl{evaluation} we believe to have developed an accurate model, however, we do not claim perfect accuracy for each property. 
Rather, we acknowledge that the prediction for individual properties may be compromised due to the variety of configurations and thermodynamic conditions considered.\\

We applied our model as part of a comparative analysis of thermally and mechanically induced rippling in graphene and hBN.
In the course of this study, we found that the bending rigidities predicted by both GAP models are significantly larger than reported by previous work based on established force fields. 
\hl{Both models predict an almost identical scaling behaviour of the rippling height with $\eta \approx 0.85$ for hBN and graphene being in accordance with the theory of flexible membranes.}
%
In contrast to the scaling behaviour, however, the actual height of the out-of-plane deviations for a given system size is directly affected by the material's resistance to bending showing larger values for hBN. 
This observation also holds for mechanically deformed ripples irrespective of whether the system was exposed to compression or tension. 
Eventually, the phase transition from randomly fluctuating to soliton-like ripples occurs at roughly the same strain rate for both materials.\\

It is important, however, to point out some of the limitations and shortcomings of our potential.
Despite being computationally several orders of magnitude more efficient than DFT, the evaluation of the SOAP descriptor is demanding and the hBN-GAP is roughly three order\hl{s} of magnitude more expensive than the established hBN force fields. 
Although we extended the range of the 2B descriptor to 10~$\mathrm{\AA}$, our model is constrained by its finite cut-off and is, thus, not able to appropriately treat long-range interactions due to electrostatics or dispersion.
Further, treating charges only implicitly prevents a correct description of the phonon dispersion curves at small wavevectors.
Additionally, in reality hBN is inevitably found with defects and dopants are incorporated in the crystal's lattice \cite{Shankar2019}.
While we did not consider such systems in this study, they will be part of future extensions of our training set and GAP model.\\

Looking forward, future applications of the hBN-GAP could involve the study of how rippling is affected by dispersion interactions in bilayer and multilayer hBN. 
Another application may include treating our model as structure generator for rippled sheets to investigate the impact of the corrugation on proton transfer \cite{Hu2014,Kroes2017} and the adsorption and, potentially, permeation of small atoms and molecules with DFT \cite{Liang2017,Sun2020}.
Moreover, the hBN-GAP can be readily combined with any other force field and, thus, could be used to study the interface with fluids such as liquid water. 
In this context, a high resolution of the phonon modes, as achieved with our model, is essential as adsorbate motion on layered materials, and through nanotubes couples with phonon modes of the solid \cite{Ma2015,Ghoufi2016,Cruz-Chu2017,Marbach2018}.\\

We have made our potential as well as training and evaluation sets freely available at [\textcolor{blue}{\url{http://www.libatoms.org}}]. 
The hBN-GAP can be readily applied within the QUIP software package and used to run MD simulations in LAMMPS \cite{Plimpton1997}.

\begin{suppinfo}

The Supporting Information is available free of charge on the
ACS Publications website at DOI: .
{\renewcommand\labelitemi{}
\begin{itemize}
    \item Comparison of DFT functionals.
    \item Benchmarking of the hBN GAP including formation energies of nanotubes, the binding curve in bulk hBN, as well as the computational efficiency.
    \item Phonon dispersion curves predicted by the various force fields.
    \item Analysis of the impact of box shape and box size on the predicted rippling amplitudes.
\end{itemize}}

 \end{suppinfo}

\begin{acknowledgement}
We thank Gábor Csányi for fruitful discussions and valuable advice during the fitting process of the GAP model. We also thank Piero Gasparotto for his guidance and support in creating the sketch-map. The authors are grateful to the UK Materials and Molecular Modelling Hub for computational resources, which is partially funded by EPSRC (EP/P020194/1).  Via our membership of the UK's HEC Materials Chemistry Consortium, which is funded by EPSRC (EP/L000202, EP/R029431), this work used the ARCHER UK National Supercomputing Service (\textcolor{blue}{http://www.archer.ac.uk}). Additionally, we are also grateful for the use of the UCL Grace High Performance Computing Facility (Grace@UCL), and associated support services, in the completion of this work. Similarly, we are also grateful for the use of the Imperial College Research Computing Service, DOI: 10.14469/hpc/2232. 
\end{acknowledgement}

\newpage


\providecommand{\latin}[1]{#1}
\makeatletter
\providecommand{\doi}
  {\begingroup\let\do\@makeother\dospecials
  \catcode`\{=1 \catcode`\}=2 \doi@aux}
\providecommand{\doi@aux}[1]{\endgroup\texttt{#1}}
\makeatother
\providecommand*\mcitethebibliography{\thebibliography}
\csname @ifundefined\endcsname{endmcitethebibliography}
  {\let\endmcitethebibliography\endthebibliography}{}
\begin{mcitethebibliography}{132}
\providecommand*\natexlab[1]{#1}
\providecommand*\mciteSetBstSublistMode[1]{}
\providecommand*\mciteSetBstMaxWidthForm[2]{}
\providecommand*\mciteBstWouldAddEndPuncttrue
  {\def\EndOfBibitem{\unskip.}}
\providecommand*\mciteBstWouldAddEndPunctfalse
  {\let\EndOfBibitem\relax}
\providecommand*\mciteSetBstMidEndSepPunct[3]{}
\providecommand*\mciteSetBstSublistLabelBeginEnd[3]{}
\providecommand*\EndOfBibitem{}
\mciteSetBstSublistMode{f}
\mciteSetBstMaxWidthForm{subitem}{(\alph{mcitesubitemcount})}
\mciteSetBstSublistLabelBeginEnd
  {\mcitemaxwidthsubitemform\space}
  {\relax}
  {\relax}

\bibitem[Watanabe \latin{et~al.}(2004)Watanabe, Taniguchi, and
  Kanda]{Watanabe2004}
Watanabe,~K.; Taniguchi,~T.; Kanda,~H. {Direct-bandgap properties and evidence
  for ultraviolet lasing of hexagonal boron nitride single crystal}.
  \emph{Nature Materials} \textbf{2004}, \emph{3}, 404--409\relax
\mciteBstWouldAddEndPuncttrue
\mciteSetBstMidEndSepPunct{\mcitedefaultmidpunct}
{\mcitedefaultendpunct}{\mcitedefaultseppunct}\relax
\EndOfBibitem
\bibitem[Chen \latin{et~al.}(2004)Chen, Zou, Campbell, and Caer]{Chen2004}
Chen,~Y.; Zou,~J.; Campbell,~S.~J.; Caer,~G.~L. {Boron nitride nanotubes:
  Pronounced resistance to oxidation}. \emph{Applied Physics Letters}
  \textbf{2004}, \emph{84}, 2430--2432\relax
\mciteBstWouldAddEndPuncttrue
\mciteSetBstMidEndSepPunct{\mcitedefaultmidpunct}
{\mcitedefaultendpunct}{\mcitedefaultseppunct}\relax
\EndOfBibitem
\bibitem[Pakdel \latin{et~al.}(2014)Pakdel, Bando, and Golberg]{Pakdel2014}
Pakdel,~A.; Bando,~Y.; Golberg,~D. {Nano boron nitride flatland}.
  \emph{Chemical Society Reviews} \textbf{2014}, \emph{43}, 934--959\relax
\mciteBstWouldAddEndPuncttrue
\mciteSetBstMidEndSepPunct{\mcitedefaultmidpunct}
{\mcitedefaultendpunct}{\mcitedefaultseppunct}\relax
\EndOfBibitem
\bibitem[Geim and Grigorieva(2013)Geim, and Grigorieva]{Geim2013}
Geim,~A.~K.; Grigorieva,~I.~V. {Van der Waals heterostructures}. \emph{Nature}
  \textbf{2013}, \emph{499}, 419--425\relax
\mciteBstWouldAddEndPuncttrue
\mciteSetBstMidEndSepPunct{\mcitedefaultmidpunct}
{\mcitedefaultendpunct}{\mcitedefaultseppunct}\relax
\EndOfBibitem
\bibitem[Britnell \latin{et~al.}(2012)Britnell, Gorbachev, Jalil, Belle,
  Schedin, Mishchenko, Georgiou, Katsnelson, Eaves, Morozov, Peres, Leist,
  Geim, Novoselov, and Ponomarenko]{Britnell2012}
Britnell,~L.; Gorbachev,~R.~V.; Jalil,~R.; Belle,~B.~D.; Schedin,~F.;
  Mishchenko,~A.; Georgiou,~T.; Katsnelson,~M.~I.; Eaves,~L.; Morozov,~S.~V.
  \latin{et~al.}  {Field-Effect Tunneling Transistor Based on Vertical Graphene
  Heterostructures}. \emph{Science} \textbf{2012}, \emph{335}, 947--951\relax
\mciteBstWouldAddEndPuncttrue
\mciteSetBstMidEndSepPunct{\mcitedefaultmidpunct}
{\mcitedefaultendpunct}{\mcitedefaultseppunct}\relax
\EndOfBibitem
\bibitem[Xia \latin{et~al.}(2014)Xia, Wang, Xiao, Dubey, and
  Ramasubramaniam]{Xia2014a}
Xia,~F.; Wang,~H.; Xiao,~D.; Dubey,~M.; Ramasubramaniam,~A. {Two-dimensional
  material nanophotonics}. \emph{Nature Photonics} \textbf{2014}, \emph{8},
  899--907\relax
\mciteBstWouldAddEndPuncttrue
\mciteSetBstMidEndSepPunct{\mcitedefaultmidpunct}
{\mcitedefaultendpunct}{\mcitedefaultseppunct}\relax
\EndOfBibitem
\bibitem[Lei \latin{et~al.}(2013)Lei, Portehault, Liu, Qin, and Chen]{Lei2013}
Lei,~W.; Portehault,~D.; Liu,~D.; Qin,~S.; Chen,~Y. {Porous boron nitride
  nanosheets for effective water cleaning}. \emph{Nature Communications}
  \textbf{2013}, \emph{4}, 1777\relax
\mciteBstWouldAddEndPuncttrue
\mciteSetBstMidEndSepPunct{\mcitedefaultmidpunct}
{\mcitedefaultendpunct}{\mcitedefaultseppunct}\relax
\EndOfBibitem
\bibitem[Zhi \latin{et~al.}(2011)Zhi, Xu, Bando, and Golberg]{Zhi2011}
Zhi,~C.; Xu,~Y.; Bando,~Y.; Golberg,~D. {Highly thermo-conductive fluid with
  boron nitride nanofillers}. \emph{ACS Nano} \textbf{2011}, \emph{5},
  6571--6577\relax
\mciteBstWouldAddEndPuncttrue
\mciteSetBstMidEndSepPunct{\mcitedefaultmidpunct}
{\mcitedefaultendpunct}{\mcitedefaultseppunct}\relax
\EndOfBibitem
\bibitem[Taha-Tijerina \latin{et~al.}(2012)Taha-Tijerina, Narayanan, Gao,
  Rohde, Tsentalovich, Pasquali, and Ajayan]{Taha-Tijerina2012}
Taha-Tijerina,~J.; Narayanan,~T.~N.; Gao,~G.; Rohde,~M.; Tsentalovich,~D.~A.;
  Pasquali,~M.; Ajayan,~P.~M. {Electrically insulating thermal nano-oils using
  2D fillers}. \emph{ACS Nano} \textbf{2012}, \emph{6}, 1214--1220\relax
\mciteBstWouldAddEndPuncttrue
\mciteSetBstMidEndSepPunct{\mcitedefaultmidpunct}
{\mcitedefaultendpunct}{\mcitedefaultseppunct}\relax
\EndOfBibitem
\bibitem[Krishnam \latin{et~al.}(2016)Krishnam, Bose, and Das]{Krishnam2016}
Krishnam,~M.; Bose,~S.; Das,~C. {Boron nitride (BN) nanofluids as cooling agent
  in thermal management system (TMS)}. \emph{Applied Thermal Engineering}
  \textbf{2016}, \emph{106}, 951--958\relax
\mciteBstWouldAddEndPuncttrue
\mciteSetBstMidEndSepPunct{\mcitedefaultmidpunct}
{\mcitedefaultendpunct}{\mcitedefaultseppunct}\relax
\EndOfBibitem
\bibitem[Zhao \latin{et~al.}(2019)Zhao, Ding, Xu, Zhao, Zheng, Shao, and
  Yu]{Zhao2019}
Zhao,~H.; Ding,~J.; Xu,~B.; Zhao,~X.; Zheng,~Y.; Shao,~Z.; Yu,~H. {Green
  Synthesis of Graphene/Boron Nitride Composites for Ultrahigh Thermally
  Conductive Fluids}. \emph{ACS Sustainable Chemistry and Engineering}
  \textbf{2019}, \emph{7}, 14266--14272\relax
\mciteBstWouldAddEndPuncttrue
\mciteSetBstMidEndSepPunct{\mcitedefaultmidpunct}
{\mcitedefaultendpunct}{\mcitedefaultseppunct}\relax
\EndOfBibitem
\bibitem[Ciofani \latin{et~al.}(2013)Ciofani, Danti, Genchi, Mazzolai, and
  Mattoli]{Ciofani2013}
Ciofani,~G.; Danti,~S.; Genchi,~G.~G.; Mazzolai,~B.; Mattoli,~V. {Boron nitride
  nanotubes: Biocompatibility and potential spill-over in nanomedicine}.
  \emph{Small} \textbf{2013}, \emph{9}, 1672--1685\relax
\mciteBstWouldAddEndPuncttrue
\mciteSetBstMidEndSepPunct{\mcitedefaultmidpunct}
{\mcitedefaultendpunct}{\mcitedefaultseppunct}\relax
\EndOfBibitem
\bibitem[Weng \latin{et~al.}(2014)Weng, Wang, Wang, Hanagata, Li, Liu, Wang,
  Jiang, Bando, and Golberg]{Weng2014}
Weng,~Q.; Wang,~B.; Wang,~X.; Hanagata,~N.; Li,~X.; Liu,~D.; Wang,~X.;
  Jiang,~X.; Bando,~Y.; Golberg,~D. {Highly water-soluble, porous, and
  biocompatible boron nitrides for anticancer drug delivery}. \emph{ACS Nano}
  \textbf{2014}, \emph{8}, 6123--6130\relax
\mciteBstWouldAddEndPuncttrue
\mciteSetBstMidEndSepPunct{\mcitedefaultmidpunct}
{\mcitedefaultendpunct}{\mcitedefaultseppunct}\relax
\EndOfBibitem
\bibitem[Weng \latin{et~al.}(2016)Weng, Wang, Wang, Bando, and
  Golberg]{Weng2016}
Weng,~Q.; Wang,~X.; Wang,~X.; Bando,~Y.; Golberg,~D. {Functionalized hexagonal
  boron nitride nanomaterials: Emerging properties and applications}.
  \emph{Chemical Society Reviews} \textbf{2016}, \emph{45}, 3989--4012\relax
\mciteBstWouldAddEndPuncttrue
\mciteSetBstMidEndSepPunct{\mcitedefaultmidpunct}
{\mcitedefaultendpunct}{\mcitedefaultseppunct}\relax
\EndOfBibitem
\bibitem[Secchi \latin{et~al.}(2016)Secchi, Marbach, Nigu{\`{e}}s, Stein,
  Siria, and Bocquet]{Secchi2016}
Secchi,~E.; Marbach,~S.; Nigu{\`{e}}s,~A.; Stein,~D.; Siria,~A.; Bocquet,~L.
  {Massive radius-dependent flow slippage in carbon nanotubes}. \emph{Nature}
  \textbf{2016}, \emph{537}, 210--213\relax
\mciteBstWouldAddEndPuncttrue
\mciteSetBstMidEndSepPunct{\mcitedefaultmidpunct}
{\mcitedefaultendpunct}{\mcitedefaultseppunct}\relax
\EndOfBibitem
\bibitem[Siria \latin{et~al.}(2013)Siria, Poncharal, Biance, Fulcrand, Blase,
  Purcell, and Bocquet]{Siria2013}
Siria,~A.; Poncharal,~P.; Biance,~A.~L.; Fulcrand,~R.; Blase,~X.;
  Purcell,~S.~T.; Bocquet,~L. {Giant osmotic energy conversion measured in a
  single transmembrane boron nitride nanotube}. \emph{Nature} \textbf{2013},
  \emph{494}, 455--458\relax
\mciteBstWouldAddEndPuncttrue
\mciteSetBstMidEndSepPunct{\mcitedefaultmidpunct}
{\mcitedefaultendpunct}{\mcitedefaultseppunct}\relax
\EndOfBibitem
\bibitem[Grosjean \latin{et~al.}(2016)Grosjean, Pean, Siria, Bocquet,
  Vuilleumier, and Bocquet]{Grosjean2016}
Grosjean,~B.; Pean,~C.; Siria,~A.; Bocquet,~L.; Vuilleumier,~R.; Bocquet,~M.~L.
  {Chemisorption of Hydroxide on 2D Materials from DFT Calculations: Graphene
  versus Hexagonal Boron Nitride}. \emph{Journal of Physical Chemistry Letters}
  \textbf{2016}, \emph{7}, 4695--4700\relax
\mciteBstWouldAddEndPuncttrue
\mciteSetBstMidEndSepPunct{\mcitedefaultmidpunct}
{\mcitedefaultendpunct}{\mcitedefaultseppunct}\relax
\EndOfBibitem
\bibitem[Secchi \latin{et~al.}(2016)Secchi, Nigu{\`{e}}s, Jubin, Siria, and
  Bocquet]{Secchi2016a}
Secchi,~E.; Nigu{\`{e}}s,~A.; Jubin,~L.; Siria,~A.; Bocquet,~L. {Scaling
  behavior for ionic transport and its fluctuations in individual carbon
  nanotubes}. \emph{Physical Review Letters} \textbf{2016}, \emph{116},
  154501\relax
\mciteBstWouldAddEndPuncttrue
\mciteSetBstMidEndSepPunct{\mcitedefaultmidpunct}
{\mcitedefaultendpunct}{\mcitedefaultseppunct}\relax
\EndOfBibitem
\bibitem[Siria \latin{et~al.}(2017)Siria, Bocquet, and Bocquet]{Siria2017}
Siria,~A.; Bocquet,~M.-L.; Bocquet,~L. {New avenues for the large-scale
  harvesting of blue energy}. \emph{Nature Reviews Chemistry} \textbf{2017},
  \emph{1}, 0091\relax
\mciteBstWouldAddEndPuncttrue
\mciteSetBstMidEndSepPunct{\mcitedefaultmidpunct}
{\mcitedefaultendpunct}{\mcitedefaultseppunct}\relax
\EndOfBibitem
\bibitem[Meyer \latin{et~al.}(2007)Meyer, Geim, Katsnelson, Novoselov, Booth,
  and Roth]{Meyer2007}
Meyer,~J.~C.; Geim,~A.~K.; Katsnelson,~M.~I.; Novoselov,~K.~S.; Booth,~T.~J.;
  Roth,~S. {The structure of suspended graphene sheets}. \emph{Nature}
  \textbf{2007}, \emph{446}, 60--63\relax
\mciteBstWouldAddEndPuncttrue
\mciteSetBstMidEndSepPunct{\mcitedefaultmidpunct}
{\mcitedefaultendpunct}{\mcitedefaultseppunct}\relax
\EndOfBibitem
\bibitem[Fasolino \latin{et~al.}(2007)Fasolino, Los, and
  Katsnelson]{Fasolino2007}
Fasolino,~A.; Los,~J.~H.; Katsnelson,~M.~I. {Intrinsic ripples in graphene}.
  \emph{Nature Materials} \textbf{2007}, \emph{6}, 858--861\relax
\mciteBstWouldAddEndPuncttrue
\mciteSetBstMidEndSepPunct{\mcitedefaultmidpunct}
{\mcitedefaultendpunct}{\mcitedefaultseppunct}\relax
\EndOfBibitem
\bibitem[Guinea \latin{et~al.}(2008)Guinea, Horovitz, and {Le
  Doussal}]{Guinea2008}
Guinea,~F.; Horovitz,~B.; {Le Doussal},~P. {Gauge field induced by ripples in
  graphene}. \emph{Physical Review B - Condensed Matter and Materials Physics}
  \textbf{2008}, \emph{77}, 205421\relax
\mciteBstWouldAddEndPuncttrue
\mciteSetBstMidEndSepPunct{\mcitedefaultmidpunct}
{\mcitedefaultendpunct}{\mcitedefaultseppunct}\relax
\EndOfBibitem
\bibitem[Guinea \latin{et~al.}(2008)Guinea, Katsnelson, and
  Vozmediano]{Guinea2008a}
Guinea,~F.; Katsnelson,~M.~I.; Vozmediano,~M.~A. {Midgap states and charge
  inhomogeneities in corrugated graphene}. \emph{Physical Review B - Condensed
  Matter and Materials Physics} \textbf{2008}, \emph{77}, 075422\relax
\mciteBstWouldAddEndPuncttrue
\mciteSetBstMidEndSepPunct{\mcitedefaultmidpunct}
{\mcitedefaultendpunct}{\mcitedefaultseppunct}\relax
\EndOfBibitem
\bibitem[{V{\'{a}}zquez De Parga} \latin{et~al.}(2008){V{\'{a}}zquez De Parga},
  Calleja, Borca, Passeggi, Hinarejos, Guinea, and Miranda]{VazquezDeParga2008}
{V{\'{a}}zquez De Parga},~A.~L.; Calleja,~F.; Borca,~B.; Passeggi,~M.~C.;
  Hinarejos,~J.~J.; Guinea,~F.; Miranda,~R. {Periodically rippled graphene:
  Growth and spatially resolved electronic structure}. \emph{Physical Review
  Letters} \textbf{2008}, \emph{100}, 056807\relax
\mciteBstWouldAddEndPuncttrue
\mciteSetBstMidEndSepPunct{\mcitedefaultmidpunct}
{\mcitedefaultendpunct}{\mcitedefaultseppunct}\relax
\EndOfBibitem
\bibitem[Boukhvalov and Katsnelson(2009)Boukhvalov, and
  Katsnelson]{Boukhvalov2009}
Boukhvalov,~D.~W.; Katsnelson,~M.~I. {Enhancement of chemical activity in
  corrugated graphene}. \emph{Journal of Physical Chemistry C} \textbf{2009},
  \emph{113}, 14176--14178\relax
\mciteBstWouldAddEndPuncttrue
\mciteSetBstMidEndSepPunct{\mcitedefaultmidpunct}
{\mcitedefaultendpunct}{\mcitedefaultseppunct}\relax
\EndOfBibitem
\bibitem[Bao \latin{et~al.}(2009)Bao, Miao, Chen, Zhang, Jang, Dames, and
  Lau]{Bao2009}
Bao,~W.; Miao,~F.; Chen,~Z.; Zhang,~H.; Jang,~W.; Dames,~C.; Lau,~C.~N.
  {Controlled ripple texturing of suspended graphene and ultrathin graphite
  membranes}. \emph{Nature Nanotechnology} \textbf{2009}, \emph{4},
  562--566\relax
\mciteBstWouldAddEndPuncttrue
\mciteSetBstMidEndSepPunct{\mcitedefaultmidpunct}
{\mcitedefaultendpunct}{\mcitedefaultseppunct}\relax
\EndOfBibitem
\bibitem[Ma \latin{et~al.}(2015)Ma, Grey, Shen, Urbakh, Wu, Liu, Liu, and
  Zheng]{Ma2015}
Ma,~M.; Grey,~F.; Shen,~L.; Urbakh,~M.; Wu,~S.; Liu,~J.~Z.; Liu,~Y.; Zheng,~Q.
  {Water transport inside carbon nanotubes mediated by phonon-induced
  oscillating friction}. \emph{Nature Nanotechnology} \textbf{2015}, \emph{10},
  692--695\relax
\mciteBstWouldAddEndPuncttrue
\mciteSetBstMidEndSepPunct{\mcitedefaultmidpunct}
{\mcitedefaultendpunct}{\mcitedefaultseppunct}\relax
\EndOfBibitem
\bibitem[Ghoufi \latin{et~al.}(2016)Ghoufi, Szymczyk, and Malfreyt]{Ghoufi2016}
Ghoufi,~A.; Szymczyk,~A.; Malfreyt,~P. {Ultrafast diffusion of Ionic Liquids
  Confined in Carbon Nanotubes}. \emph{Scientific Reports} \textbf{2016},
  \emph{6}, 28518\relax
\mciteBstWouldAddEndPuncttrue
\mciteSetBstMidEndSepPunct{\mcitedefaultmidpunct}
{\mcitedefaultendpunct}{\mcitedefaultseppunct}\relax
\EndOfBibitem
\bibitem[Cruz-Ch{\'{u}} \latin{et~al.}(2017)Cruz-Ch{\'{u}}, Papadopoulou,
  Walther, Popadi{\'{c}}, Li, Praprotnik, and Koumoutsakos]{Cruz-Chu2017}
Cruz-Ch{\'{u}},~E.~R.; Papadopoulou,~E.; Walther,~J.~H.; Popadi{\'{c}},~A.;
  Li,~G.; Praprotnik,~M.; Koumoutsakos,~P. {On phonons and water flow
  enhancement in carbon nanotubes}. \emph{Nature Nanotechnology} \textbf{2017},
  \emph{12}, 1106--1108\relax
\mciteBstWouldAddEndPuncttrue
\mciteSetBstMidEndSepPunct{\mcitedefaultmidpunct}
{\mcitedefaultendpunct}{\mcitedefaultseppunct}\relax
\EndOfBibitem
\bibitem[Marbach \latin{et~al.}(2018)Marbach, Dean, and Bocquet]{Marbach2018}
Marbach,~S.; Dean,~D.~S.; Bocquet,~L. {Transport and dispersion across wiggling
  nanopores}. \emph{Nature Physics} \textbf{2018}, \emph{14}, 1108--1113\relax
\mciteBstWouldAddEndPuncttrue
\mciteSetBstMidEndSepPunct{\mcitedefaultmidpunct}
{\mcitedefaultendpunct}{\mcitedefaultseppunct}\relax
\EndOfBibitem
\bibitem[Tersoff(1986)]{Tersoff1986}
Tersoff,~J. {New Empirical Model for the Structural Properties of Silicon}.
  \emph{Physical Review Letters} \textbf{1986}, \emph{56}, 632--635\relax
\mciteBstWouldAddEndPuncttrue
\mciteSetBstMidEndSepPunct{\mcitedefaultmidpunct}
{\mcitedefaultendpunct}{\mcitedefaultseppunct}\relax
\EndOfBibitem
\bibitem[Tersoff(1988)]{Tersoff1988}
Tersoff,~J. {Empirical Interatomic Potential for Carbon, with Applications to
  Amorphous Carbon}. \emph{Physical Review Letters} \textbf{1988}, \emph{61},
  2879--2882\relax
\mciteBstWouldAddEndPuncttrue
\mciteSetBstMidEndSepPunct{\mcitedefaultmidpunct}
{\mcitedefaultendpunct}{\mcitedefaultseppunct}\relax
\EndOfBibitem
\bibitem[Tersoff(1989)]{Tersoff1989}
Tersoff,~J. {Modeling solid-state chemistry: Interatomic potentials for
  multicomponent systems}. \emph{Physical Review B} \textbf{1989}, \emph{39},
  5566--5568\relax
\mciteBstWouldAddEndPuncttrue
\mciteSetBstMidEndSepPunct{\mcitedefaultmidpunct}
{\mcitedefaultendpunct}{\mcitedefaultseppunct}\relax
\EndOfBibitem
\bibitem[Brenner(1990)]{Brenner1990}
Brenner,~D.~W. {Empirical potential for hydrocarbons for use in simulating the
  chemical vapor deposition of diamond film}. \emph{Physical Review B}
  \textbf{1990}, \emph{42}, 9458--9471\relax
\mciteBstWouldAddEndPuncttrue
\mciteSetBstMidEndSepPunct{\mcitedefaultmidpunct}
{\mcitedefaultendpunct}{\mcitedefaultseppunct}\relax
\EndOfBibitem
\bibitem[Albe \latin{et~al.}(1997)Albe, M{\"{o}}ller, and Heinig]{Albe1997}
Albe,~K.; M{\"{o}}ller,~W.; Heinig,~K.~H. {Computer simulation and boron
  nitride}. \emph{Radiation Effects and Defects in Solids} \textbf{1997},
  \emph{141}, 85--97\relax
\mciteBstWouldAddEndPuncttrue
\mciteSetBstMidEndSepPunct{\mcitedefaultmidpunct}
{\mcitedefaultendpunct}{\mcitedefaultseppunct}\relax
\EndOfBibitem
\bibitem[Sekkal \latin{et~al.}(1998)Sekkal, Bouhafs, Aourag, and
  Certier]{Sekkal1998}
Sekkal,~W.; Bouhafs,~B.; Aourag,~H.; Certier,~M. {Molecular-dynamics simulation
  of structural and thermodynamic properties of boron nitride}. \emph{Journal
  of Physics Condensed Matter} \textbf{1998}, \emph{10}, 4975--4984\relax
\mciteBstWouldAddEndPuncttrue
\mciteSetBstMidEndSepPunct{\mcitedefaultmidpunct}
{\mcitedefaultendpunct}{\mcitedefaultseppunct}\relax
\EndOfBibitem
\bibitem[Matsunaga \latin{et~al.}(2000)Matsunaga, Fisher, and
  Matsubara]{Matsunaga2000}
Matsunaga,~K.; Fisher,~C.; Matsubara,~H. {Tersoff Potential Parameters for
  Simulating Cubic Boron Carbonitrides solid solution model Tersoff Potential
  Parameters for Simulating Cubic Boron Carbonitrides}. \emph{Japanese Journal
  of Applied Physics} \textbf{2000}, \emph{39}, 48--51\relax
\mciteBstWouldAddEndPuncttrue
\mciteSetBstMidEndSepPunct{\mcitedefaultmidpunct}
{\mcitedefaultendpunct}{\mcitedefaultseppunct}\relax
\EndOfBibitem
\bibitem[Matsunaga and Iwamoto(2001)Matsunaga, and Iwamoto]{Matsunaga2001}
Matsunaga,~K.; Iwamoto,~Y. {Molecular Dynamics Study of Atomic Structure and
  Diffusion Behavior in Amorphous Silicon Nitride Containing Boron}.
  \emph{Journal of the American Ceramic Society} \textbf{2001}, \emph{84},
  2213--2219\relax
\mciteBstWouldAddEndPuncttrue
\mciteSetBstMidEndSepPunct{\mcitedefaultmidpunct}
{\mcitedefaultendpunct}{\mcitedefaultseppunct}\relax
\EndOfBibitem
\bibitem[Sevik \latin{et~al.}(2011)Sevik, Kınacı, Haskins, and
  {\c{C}}ağın]{Sevik2011}
Sevik,~C.; Kınacı,~A.; Haskins,~J.~B.; {\c{C}}ağın,~T. {Characterization of
  thermal transport in low-dimensional boron nitride nanostructures}.
  \emph{Physical Review B - Condensed Matter and Materials Physics}
  \textbf{2011}, \emph{84}, 085409\relax
\mciteBstWouldAddEndPuncttrue
\mciteSetBstMidEndSepPunct{\mcitedefaultmidpunct}
{\mcitedefaultendpunct}{\mcitedefaultseppunct}\relax
\EndOfBibitem
\bibitem[Kınacı \latin{et~al.}(2012)Kınacı, Haskins, Sevik, and
  {\c{C}}ağın]{Kinaci2012}
Kınacı,~A.; Haskins,~J.~B.; Sevik,~C.; {\c{C}}ağın,~T. {Thermal
  conductivity of BN-C nanostructures}. \emph{Physical Review B - Condensed
  Matter and Materials Physics} \textbf{2012}, \emph{86}, 115410\relax
\mciteBstWouldAddEndPuncttrue
\mciteSetBstMidEndSepPunct{\mcitedefaultmidpunct}
{\mcitedefaultendpunct}{\mcitedefaultseppunct}\relax
\EndOfBibitem
\bibitem[Los \latin{et~al.}(2017)Los, Kroes, Albe, Gordillo, Katsnelson, and
  Fasolino]{Los2017}
Los,~J.~H.; Kroes,~J.~M.; Albe,~K.; Gordillo,~R.~M.; Katsnelson,~M.~I.;
  Fasolino,~A. {Extended Tersoff potential for boron nitride: Energetics and
  elastic properties of pristine and defective h -BN}. \emph{Physical Review B}
  \textbf{2017}, \emph{96}, 184108\relax
\mciteBstWouldAddEndPuncttrue
\mciteSetBstMidEndSepPunct{\mcitedefaultmidpunct}
{\mcitedefaultendpunct}{\mcitedefaultseppunct}\relax
\EndOfBibitem
\bibitem[van Duin \latin{et~al.}(2001)van Duin, Dasgupta, Lorant, and {Goddard
  III}]{VanDuin2001}
van Duin,~A. C.~T.; Dasgupta,~S.; Lorant,~F.; {Goddard III},~W.~A. {ReaxFF: A
  Reactive Force Field for Hydrocarbons}. \emph{Journal of Physical Chemistry
  A} \textbf{2001}, \emph{105}, 9396--9409\relax
\mciteBstWouldAddEndPuncttrue
\mciteSetBstMidEndSepPunct{\mcitedefaultmidpunct}
{\mcitedefaultendpunct}{\mcitedefaultseppunct}\relax
\EndOfBibitem
\bibitem[Chenoweth \latin{et~al.}(2008)Chenoweth, {Van Duin}, and
  Goddard]{Chenoweth2008}
Chenoweth,~K.; {Van Duin},~A.~C.; Goddard,~W.~A. {ReaxFF reactive force field
  for molecular dynamics simulations of hydrocarbon oxidation}. \emph{Journal
  of Physical Chemistry A} \textbf{2008}, \emph{112}, 1040--1053\relax
\mciteBstWouldAddEndPuncttrue
\mciteSetBstMidEndSepPunct{\mcitedefaultmidpunct}
{\mcitedefaultendpunct}{\mcitedefaultseppunct}\relax
\EndOfBibitem
\bibitem[Weismiller \latin{et~al.}(2010)Weismiller, Duin, Lee, and
  Yetter]{Weismiller2010}
Weismiller,~M.~R.; Duin,~A.~C.; Lee,~J.; Yetter,~R.~A. {ReaxFF reactive force
  field development and applications for molecular dynamics simulations of
  ammonia borane dehydrogenation and combustion}. \emph{Journal of Physical
  Chemistry A} \textbf{2010}, \emph{114}, 5485--5492\relax
\mciteBstWouldAddEndPuncttrue
\mciteSetBstMidEndSepPunct{\mcitedefaultmidpunct}
{\mcitedefaultendpunct}{\mcitedefaultseppunct}\relax
\EndOfBibitem
\bibitem[Paupitz \latin{et~al.}(2014)Paupitz, Junkermeier, van Duin, and
  Branicio]{Paupitz2014}
Paupitz,~R.; Junkermeier,~C.~E.; van Duin,~A.~C.; Branicio,~P.~S. {Fullerenes
  generated from porous structures}. \emph{Physical Chemistry Chemical Physics}
  \textbf{2014}, \emph{16}, 25515--25522\relax
\mciteBstWouldAddEndPuncttrue
\mciteSetBstMidEndSepPunct{\mcitedefaultmidpunct}
{\mcitedefaultendpunct}{\mcitedefaultseppunct}\relax
\EndOfBibitem
\bibitem[Liu \latin{et~al.}(2017)Liu, {Van Duin}, {Van Duin}, Liu, and
  Edgar]{Liu2017}
Liu,~S.; {Van Duin},~A.~C.; {Van Duin},~D.~M.; Liu,~B.; Edgar,~J.~H. {Atomistic
  Insights into Nucleation and Formation of Hexagonal Boron Nitride on Nickel
  from First-Principles-Based Reactive Molecular Dynamics Simulations}.
  \emph{ACS Nano} \textbf{2017}, \emph{11}, 3585--3596\relax
\mciteBstWouldAddEndPuncttrue
\mciteSetBstMidEndSepPunct{\mcitedefaultmidpunct}
{\mcitedefaultendpunct}{\mcitedefaultseppunct}\relax
\EndOfBibitem
\bibitem[Rajan \latin{et~al.}(2018)Rajan, Strano, and Blankschtein]{Rajan2018}
Rajan,~A.~G.; Strano,~M.~S.; Blankschtein,~D. {Ab Initio Molecular Dynamics and
  Lattice Dynamics-Based Force Field for Modeling Hexagonal Boron Nitride in
  Mechanical and Interfacial Applications}. \emph{Journal of Physical Chemistry
  Letters} \textbf{2018}, \emph{9}, 1584--1591\relax
\mciteBstWouldAddEndPuncttrue
\mciteSetBstMidEndSepPunct{\mcitedefaultmidpunct}
{\mcitedefaultendpunct}{\mcitedefaultseppunct}\relax
\EndOfBibitem
\bibitem[Mortazavi and R{\'{e}}mond(2012)Mortazavi, and
  R{\'{e}}mond]{Mortazavi2012}
Mortazavi,~B.; R{\'{e}}mond,~Y. {Investigation of tensile response and thermal
  conductivity of boron-nitride nanosheets using molecular dynamics
  simulations}. \emph{Physica E: Low-Dimensional Systems and Nanostructures}
  \textbf{2012}, \emph{44}, 1846--1852\relax
\mciteBstWouldAddEndPuncttrue
\mciteSetBstMidEndSepPunct{\mcitedefaultmidpunct}
{\mcitedefaultendpunct}{\mcitedefaultseppunct}\relax
\EndOfBibitem
\bibitem[Liao \latin{et~al.}(2012)Liao, Lian, and Ju]{Liao2012}
Liao,~M.-L.; Lian,~T.-W.; Ju,~S.-P. {Tensile and compressive behaviours of a
  boron nitride nanotube: Temperature effects}. \emph{Materials Science Forum}
  \textbf{2012}, \emph{700}, 125--128\relax
\mciteBstWouldAddEndPuncttrue
\mciteSetBstMidEndSepPunct{\mcitedefaultmidpunct}
{\mcitedefaultendpunct}{\mcitedefaultseppunct}\relax
\EndOfBibitem
\bibitem[Krishnan and Ghosh(2014)Krishnan, and Ghosh]{AnoopKrishnan2014}
Krishnan,~N. M.~A.; Ghosh,~D. {Defect induced plasticity and failure mechanism
  of boron nitride nanotubes under tension}. \emph{Journal of Applied Physics}
  \textbf{2014}, \emph{116}, 044313\relax
\mciteBstWouldAddEndPuncttrue
\mciteSetBstMidEndSepPunct{\mcitedefaultmidpunct}
{\mcitedefaultendpunct}{\mcitedefaultseppunct}\relax
\EndOfBibitem
\bibitem[Le and Nguyen(2014)Le, and Nguyen]{Le2014}
Le,~M.~Q.; Nguyen,~D.~T. {Atomistic simulations of pristine and defective
  hexagonal BN and SiC sheets under uniaxial tension}. \emph{Materials Science
  and Engineering A} \textbf{2014}, \emph{615}, 481--488\relax
\mciteBstWouldAddEndPuncttrue
\mciteSetBstMidEndSepPunct{\mcitedefaultmidpunct}
{\mcitedefaultendpunct}{\mcitedefaultseppunct}\relax
\EndOfBibitem
\bibitem[Hu \latin{et~al.}(2014)Hu, Lozada-Hidalgo, Wang, Mishchenko, Schedin,
  Nair, Hill, Boukhvalov, Katsnelson, Dryfe, Grigorieva, Wu, and Geim]{Hu2014}
Hu,~S.; Lozada-Hidalgo,~M.; Wang,~F.~C.; Mishchenko,~A.; Schedin,~F.;
  Nair,~R.~R.; Hill,~E.~W.; Boukhvalov,~D.~W.; Katsnelson,~M.~I.; Dryfe,~R.~A.
  \latin{et~al.}  {Proton transport through one-atom-thick crystals}.
  \emph{Nature} \textbf{2014}, \emph{516}, 227--230\relax
\mciteBstWouldAddEndPuncttrue
\mciteSetBstMidEndSepPunct{\mcitedefaultmidpunct}
{\mcitedefaultendpunct}{\mcitedefaultseppunct}\relax
\EndOfBibitem
\bibitem[Tocci \latin{et~al.}(2014)Tocci, Joly, and Michaelides]{Tocci2014a}
Tocci,~G.; Joly,~L.; Michaelides,~A. {Friction of water on graphene and
  hexagonal boron nitride from Ab initio methods: Very different slippage
  despite very similar interface structures}. \emph{Nano Letters}
  \textbf{2014}, \emph{14}, 6872--6877\relax
\mciteBstWouldAddEndPuncttrue
\mciteSetBstMidEndSepPunct{\mcitedefaultmidpunct}
{\mcitedefaultendpunct}{\mcitedefaultseppunct}\relax
\EndOfBibitem
\bibitem[Androulidakis \latin{et~al.}(2018)Androulidakis, Koukaras, Poss,
  Papagelis, Galiotis, and Tawfick]{Androulidakis2018}
Androulidakis,~C.; Koukaras,~E.~N.; Poss,~M.; Papagelis,~K.; Galiotis,~C.;
  Tawfick,~S. {Strained hexagonal boron nitride: Phonon shift and
  Gr{\"{u}}neisen parameter}. \emph{Physical Review B} \textbf{2018},
  \emph{97}, 241414\relax
\mciteBstWouldAddEndPuncttrue
\mciteSetBstMidEndSepPunct{\mcitedefaultmidpunct}
{\mcitedefaultendpunct}{\mcitedefaultseppunct}\relax
\EndOfBibitem
\bibitem[Simonnin \latin{et~al.}(2017)Simonnin, Noetinger, Nieto-Draghi, Marry,
  and Rotenberg]{Simonnin2017}
Simonnin,~P.; Noetinger,~B.; Nieto-Draghi,~C.; Marry,~V.; Rotenberg,~B.
  {Diffusion under Confinement: Hydrodynamic Finite-Size Effects in
  Simulation}. \emph{Journal of Chemical Theory and Computation} \textbf{2017},
  \emph{13}, 2881--2889\relax
\mciteBstWouldAddEndPuncttrue
\mciteSetBstMidEndSepPunct{\mcitedefaultmidpunct}
{\mcitedefaultendpunct}{\mcitedefaultseppunct}\relax
\EndOfBibitem
\bibitem[Los \latin{et~al.}(2009)Los, Katsnelson, Yazyev, Zakharchenko, and
  Fasolino]{Los2009}
Los,~J.~H.; Katsnelson,~M.~I.; Yazyev,~O.~V.; Zakharchenko,~K.~V.; Fasolino,~A.
  {Scaling properties of flexible membranes from atomistic simulations:
  Application to graphene}. \emph{Physical Review B - Condensed Matter and
  Materials Physics} \textbf{2009}, \emph{80}, 1--4\relax
\mciteBstWouldAddEndPuncttrue
\mciteSetBstMidEndSepPunct{\mcitedefaultmidpunct}
{\mcitedefaultendpunct}{\mcitedefaultseppunct}\relax
\EndOfBibitem
\bibitem[Rupp \latin{et~al.}(2012)Rupp, Tkatchenko, M{\"{u}}ller, and {Von
  Lilienfeld}]{Rupp2012}
Rupp,~M.; Tkatchenko,~A.; M{\"{u}}ller,~K.~R.; {Von Lilienfeld},~O.~A. {Fast
  and accurate modeling of molecular atomization energies with machine
  learning}. \emph{Physical Review Letters} \textbf{2012}, \emph{108},
  058301\relax
\mciteBstWouldAddEndPuncttrue
\mciteSetBstMidEndSepPunct{\mcitedefaultmidpunct}
{\mcitedefaultendpunct}{\mcitedefaultseppunct}\relax
\EndOfBibitem
\bibitem[Behler(2016)]{Behler2016}
Behler,~J. {Perspective : Machine learning potentials for atomistic
  simulations}. \emph{Journal of Chemical Physics} \textbf{2016}, \emph{145},
  170901\relax
\mciteBstWouldAddEndPuncttrue
\mciteSetBstMidEndSepPunct{\mcitedefaultmidpunct}
{\mcitedefaultendpunct}{\mcitedefaultseppunct}\relax
\EndOfBibitem
\bibitem[Bart{\'{o}}k \latin{et~al.}(2017)Bart{\'{o}}k, Cs{\'{a}}nyi, Kermode,
  De, Bernstein, Ceriotti, Bart{\'{o}}k, and Poelking]{Csanyi2017}
Bart{\'{o}}k,~A.~P.; Cs{\'{a}}nyi,~G.; Kermode,~J.~R.; De,~S.; Bernstein,~N.;
  Ceriotti,~M.; Bart{\'{o}}k,~A.~P.; Poelking,~C. {Machine learning unifies the
  modeling of materials and molecules}. \emph{Science Advances} \textbf{2017},
  \emph{3}, e1701816\relax
\mciteBstWouldAddEndPuncttrue
\mciteSetBstMidEndSepPunct{\mcitedefaultmidpunct}
{\mcitedefaultendpunct}{\mcitedefaultseppunct}\relax
\EndOfBibitem
\bibitem[Botu \latin{et~al.}(2017)Botu, Batra, Chapman, and
  Ramprasad]{Botu2017}
Botu,~V.; Batra,~R.; Chapman,~J.; Ramprasad,~R. {Machine learning force fields:
  Construction, validation, and outlook}. \emph{Journal of Physical Chemistry
  C} \textbf{2017}, \emph{121}, 511--522\relax
\mciteBstWouldAddEndPuncttrue
\mciteSetBstMidEndSepPunct{\mcitedefaultmidpunct}
{\mcitedefaultendpunct}{\mcitedefaultseppunct}\relax
\EndOfBibitem
\bibitem[Li \latin{et~al.}(2017)Li, Wang, Chin, Achenie, and Xin]{Li2017}
Li,~Z.; Wang,~S.; Chin,~W.~S.; Achenie,~L.~E.; Xin,~H. {High-throughput
  screening of bimetallic catalysts enabled by machine learning}. \emph{Journal
  of Materials Chemistry A} \textbf{2017}, \emph{5}, 24131--24138\relax
\mciteBstWouldAddEndPuncttrue
\mciteSetBstMidEndSepPunct{\mcitedefaultmidpunct}
{\mcitedefaultendpunct}{\mcitedefaultseppunct}\relax
\EndOfBibitem
\bibitem[Kitchin(2018)]{Kitchin2018}
Kitchin,~J.~R. {Machine learning in catalysis}. \emph{Nature Catalysis}
  \textbf{2018}, \emph{1}, 230--232\relax
\mciteBstWouldAddEndPuncttrue
\mciteSetBstMidEndSepPunct{\mcitedefaultmidpunct}
{\mcitedefaultendpunct}{\mcitedefaultseppunct}\relax
\EndOfBibitem
\bibitem[Goldsmith \latin{et~al.}(2018)Goldsmith, Esterhuizen, Liu, Bartel, and
  Sutton]{Goldsmith2018}
Goldsmith,~B.~R.; Esterhuizen,~J.; Liu,~J.~X.; Bartel,~C.~J.; Sutton,~C.
  {Machine learning for heterogeneous catalyst design and discovery}.
  \emph{AIChE Journal} \textbf{2018}, \emph{64}, 2311--2323\relax
\mciteBstWouldAddEndPuncttrue
\mciteSetBstMidEndSepPunct{\mcitedefaultmidpunct}
{\mcitedefaultendpunct}{\mcitedefaultseppunct}\relax
\EndOfBibitem
\bibitem[Behler and Parrinello(2007)Behler, and Parrinello]{Behler2007}
Behler,~J.; Parrinello,~M. {Generalized neural-network representation of
  high-dimensional potential-energy surfaces}. \emph{Physical Review Letters}
  \textbf{2007}, \emph{98}, 146401\relax
\mciteBstWouldAddEndPuncttrue
\mciteSetBstMidEndSepPunct{\mcitedefaultmidpunct}
{\mcitedefaultendpunct}{\mcitedefaultseppunct}\relax
\EndOfBibitem
\bibitem[Bart{\'{o}}k \latin{et~al.}(2010)Bart{\'{o}}k, Payne, Kondor, and
  Cs{\'{a}}nyi]{Bartok2010}
Bart{\'{o}}k,~A.~P.; Payne,~M.~C.; Kondor,~R.; Cs{\'{a}}nyi,~G. {Gaussian
  Approximation Potentials : The Accuracy of Quantum Mechanics , without the
  Electrons}. \emph{Physical Review Letters} \textbf{2010}, \emph{104},
  136403\relax
\mciteBstWouldAddEndPuncttrue
\mciteSetBstMidEndSepPunct{\mcitedefaultmidpunct}
{\mcitedefaultendpunct}{\mcitedefaultseppunct}\relax
\EndOfBibitem
\bibitem[Li \latin{et~al.}(2015)Li, Kermode, and {De Vita}]{Li2015}
Li,~Z.; Kermode,~J.~R.; {De Vita},~A. {Molecular Dynamics with On-the-Fly
  Machine Learning of Quantum-Mechanical Forces}. \emph{Physical Review
  Letters} \textbf{2015}, \emph{114}, 096405\relax
\mciteBstWouldAddEndPuncttrue
\mciteSetBstMidEndSepPunct{\mcitedefaultmidpunct}
{\mcitedefaultendpunct}{\mcitedefaultseppunct}\relax
\EndOfBibitem
\bibitem[Chmiela \latin{et~al.}(2017)Chmiela, Tkatchenko, Sauceda, Poltavsky,
  Sch{\"{u}}tt, and M{\"{u}}ller]{Chmiela2017a}
Chmiela,~S.; Tkatchenko,~A.; Sauceda,~H.~E.; Poltavsky,~I.;
  Sch{\"{u}}tt,~K.~T.; M{\"{u}}ller,~K.~R. {Machine learning of accurate
  energy-conserving molecular force fields}. \emph{Science Advances}
  \textbf{2017}, \emph{3}, e1603015\relax
\mciteBstWouldAddEndPuncttrue
\mciteSetBstMidEndSepPunct{\mcitedefaultmidpunct}
{\mcitedefaultendpunct}{\mcitedefaultseppunct}\relax
\EndOfBibitem
\bibitem[Nguyen \latin{et~al.}(2018)Nguyen, Sz{\'{e}}kely, Imbalzano, Behler,
  Cs{\'{a}}nyi, Ceriotti, G{\"{o}}tz, and Paesani]{Nguyen2018}
Nguyen,~T.~T.; Sz{\'{e}}kely,~E.; Imbalzano,~G.; Behler,~J.; Cs{\'{a}}nyi,~G.;
  Ceriotti,~M.; G{\"{o}}tz,~A.~W.; Paesani,~F. {Comparison of permutationally
  invariant polynomials, neural networks, and Gaussian approximation potentials
  in representing water interactions through many-body expansions}.
  \emph{Journal of Chemical Physics} \textbf{2018}, \emph{148}, 241725\relax
\mciteBstWouldAddEndPuncttrue
\mciteSetBstMidEndSepPunct{\mcitedefaultmidpunct}
{\mcitedefaultendpunct}{\mcitedefaultseppunct}\relax
\EndOfBibitem
\bibitem[Bart{\'{o}}k and Cs{\'{a}}nyi(2015)Bart{\'{o}}k, and
  Cs{\'{a}}nyi]{Bartok2015}
Bart{\'{o}}k,~A.~P.; Cs{\'{a}}nyi,~G. {Gaussian approximation potentials: A
  brief tutorial introduction}. \emph{International Journal of Quantum
  Chemistry} \textbf{2015}, \emph{115}, 1051--1057\relax
\mciteBstWouldAddEndPuncttrue
\mciteSetBstMidEndSepPunct{\mcitedefaultmidpunct}
{\mcitedefaultendpunct}{\mcitedefaultseppunct}\relax
\EndOfBibitem
\bibitem[Rowe \latin{et~al.}(2018)Rowe, Cs{\'{a}}nyi, Alf{\`{e}}, and
  Michaelides]{Rowe2018}
Rowe,~P.; Cs{\'{a}}nyi,~G.; Alf{\`{e}},~D.; Michaelides,~A. {Development of a
  machine learning potential for graphene}. \emph{Physical Review B}
  \textbf{2018}, \emph{97}, 054303\relax
\mciteBstWouldAddEndPuncttrue
\mciteSetBstMidEndSepPunct{\mcitedefaultmidpunct}
{\mcitedefaultendpunct}{\mcitedefaultseppunct}\relax
\EndOfBibitem
\bibitem[Deringer and Cs{\'{a}}nyi(2017)Deringer, and
  Cs{\'{a}}nyi]{Deringer2017}
Deringer,~V.~L.; Cs{\'{a}}nyi,~G. {Machine learning based interatomic potential
  for amorphous carbon}. \emph{Physical Review B} \textbf{2017}, \emph{95},
  094203\relax
\mciteBstWouldAddEndPuncttrue
\mciteSetBstMidEndSepPunct{\mcitedefaultmidpunct}
{\mcitedefaultendpunct}{\mcitedefaultseppunct}\relax
\EndOfBibitem
\bibitem[Rowe \latin{et~al.}(2020)Rowe, Deringer, Gasparotto, Cs{\'{a}}nyi, and
  Michaelides]{Rowe2020b}
Rowe,~P.; Deringer,~V.~L.; Gasparotto,~P.; Cs{\'{a}}nyi,~G.; Michaelides,~A.
  {An accurate and transferable machine learning potential for carbon}.
  \emph{Journal of Chemical Physics} \textbf{2020}, \emph{153}\relax
\mciteBstWouldAddEndPuncttrue
\mciteSetBstMidEndSepPunct{\mcitedefaultmidpunct}
{\mcitedefaultendpunct}{\mcitedefaultseppunct}\relax
\EndOfBibitem
\bibitem[Szlachta \latin{et~al.}(2014)Szlachta, Bart{\'{o}}k, and
  Cs{\'{a}}nyi]{Szlachta2014}
Szlachta,~W.~J.; Bart{\'{o}}k,~A.~P.; Cs{\'{a}}nyi,~G. {Accuracy and
  transferability of GAP models for tungsten}. \emph{Physical Review B}
  \textbf{2014}, \emph{90}, 104108\relax
\mciteBstWouldAddEndPuncttrue
\mciteSetBstMidEndSepPunct{\mcitedefaultmidpunct}
{\mcitedefaultendpunct}{\mcitedefaultseppunct}\relax
\EndOfBibitem
\bibitem[Jinnouchi \latin{et~al.}(2019)Jinnouchi, Lahnsteiner, Karsai, Kresse,
  and Bokdam]{Jinnouchi2019}
Jinnouchi,~R.; Lahnsteiner,~J.; Karsai,~F.; Kresse,~G.; Bokdam,~M. {Phase
  Transitions of Hybrid Perovskites Simulated by Machine-Learning Force Fields
  Trained on the Fly with Bayesian Inference}. \emph{Physical Review Letters}
  \textbf{2019}, \emph{122}, 225701\relax
\mciteBstWouldAddEndPuncttrue
\mciteSetBstMidEndSepPunct{\mcitedefaultmidpunct}
{\mcitedefaultendpunct}{\mcitedefaultseppunct}\relax
\EndOfBibitem
\bibitem[De \latin{et~al.}(2016)De, Bart{\'{o}}k, Cs{\'{a}}nyi, and
  Ceriotti]{De2016}
De,~S.; Bart{\'{o}}k,~A.~P.; Cs{\'{a}}nyi,~G.; Ceriotti,~M. {Comparing
  molecules and solids across structural and alchemical space}. \emph{Physical
  Chemistry Chemical Physics} \textbf{2016}, \emph{18}, 13754--13769\relax
\mciteBstWouldAddEndPuncttrue
\mciteSetBstMidEndSepPunct{\mcitedefaultmidpunct}
{\mcitedefaultendpunct}{\mcitedefaultseppunct}\relax
\EndOfBibitem
\bibitem[Bart{\'{o}}k \latin{et~al.}(2013)Bart{\'{o}}k, Kondor, and
  Cs{\'{a}}nyi]{Bartok2013}
Bart{\'{o}}k,~A.~P.; Kondor,~R.; Cs{\'{a}}nyi,~G. {On representing chemical
  environments}. \emph{Physical Review B - Condensed Matter and Materials
  Physics} \textbf{2013}, \emph{87}, 184115\relax
\mciteBstWouldAddEndPuncttrue
\mciteSetBstMidEndSepPunct{\mcitedefaultmidpunct}
{\mcitedefaultendpunct}{\mcitedefaultseppunct}\relax
\EndOfBibitem
\bibitem[Smith \latin{et~al.}(2018)Smith, Nebgen, Lubbers, Isayev, and
  Roitberg]{Smith2018}
Smith,~J.~S.; Nebgen,~B.; Lubbers,~N.; Isayev,~O.; Roitberg,~A.~E. {Less is
  more: Sampling chemical space with active learning}. \emph{Journal of
  Chemical Physics} \textbf{2018}, \emph{148}\relax
\mciteBstWouldAddEndPuncttrue
\mciteSetBstMidEndSepPunct{\mcitedefaultmidpunct}
{\mcitedefaultendpunct}{\mcitedefaultseppunct}\relax
\EndOfBibitem
\bibitem[Ceriotti \latin{et~al.}(2011)Ceriotti, Tribello, and
  Parrinello]{Ceriotti2011}
Ceriotti,~M.; Tribello,~G.~A.; Parrinello,~M. {Simplifying the representation
  of complex free-energy landscapes using sketch-map}. \emph{Proceedings of the
  National Academy of Sciences} \textbf{2011}, \emph{108}, 13023--13028\relax
\mciteBstWouldAddEndPuncttrue
\mciteSetBstMidEndSepPunct{\mcitedefaultmidpunct}
{\mcitedefaultendpunct}{\mcitedefaultseppunct}\relax
\EndOfBibitem
\bibitem[Tribello \latin{et~al.}(2012)Tribello, Ceriotti, and
  Parrinello]{Tribello2012}
Tribello,~G.~A.; Ceriotti,~M.; Parrinello,~M. {Using sketch-map coordinates to
  analyze and bias molecular dynamics simulations}. \emph{Proceedings of the
  National Academy of Sciences of the United States of America} \textbf{2012},
  \emph{109}, 5196--5201\relax
\mciteBstWouldAddEndPuncttrue
\mciteSetBstMidEndSepPunct{\mcitedefaultmidpunct}
{\mcitedefaultendpunct}{\mcitedefaultseppunct}\relax
\EndOfBibitem
\bibitem[Kresse and Hafner(1993)Kresse, and Hafner]{Kresse1993}
Kresse,~G.; Hafner,~J. {Ab initio molecular dynamics for liquid metals}.
  \emph{Physical Review B} \textbf{1993}, \emph{47}, 558--561\relax
\mciteBstWouldAddEndPuncttrue
\mciteSetBstMidEndSepPunct{\mcitedefaultmidpunct}
{\mcitedefaultendpunct}{\mcitedefaultseppunct}\relax
\EndOfBibitem
\bibitem[Kresse and Hafner(1994)Kresse, and Hafner]{Kresse1994}
Kresse,~G.; Hafner,~J. {Ab initio molecular-dynamics simulation of the
  liquid-metalamorphous- semiconductor transition in germanium}. \emph{Physical
  Review B} \textbf{1994}, \emph{49}, 14251--14269\relax
\mciteBstWouldAddEndPuncttrue
\mciteSetBstMidEndSepPunct{\mcitedefaultmidpunct}
{\mcitedefaultendpunct}{\mcitedefaultseppunct}\relax
\EndOfBibitem
\bibitem[Kresse and Furthm{\"{u}}ller(1996)Kresse, and
  Furthm{\"{u}}ller]{Kresse1996}
Kresse,~G.; Furthm{\"{u}}ller,~J. {Efficient iterative schemes for ab initio
  total-energy calculations using a plane-wave basis set}. \emph{Physical
  Review B - Condensed Matter and Materials Physics} \textbf{1996}, \emph{54},
  11169--11186\relax
\mciteBstWouldAddEndPuncttrue
\mciteSetBstMidEndSepPunct{\mcitedefaultmidpunct}
{\mcitedefaultendpunct}{\mcitedefaultseppunct}\relax
\EndOfBibitem
\bibitem[Kresse and Furthm{\"{u}}ller(1996)Kresse, and
  Furthm{\"{u}}ller]{Kresse1996a}
Kresse,~G.; Furthm{\"{u}}ller,~J. {Efficiency of ab-initio total energy
  calculations for metals and semiconductors using a plane-wave basis set}.
  \emph{Computational Materials Science} \textbf{1996}, \emph{6}, 15--50\relax
\mciteBstWouldAddEndPuncttrue
\mciteSetBstMidEndSepPunct{\mcitedefaultmidpunct}
{\mcitedefaultendpunct}{\mcitedefaultseppunct}\relax
\EndOfBibitem
\bibitem[Perdew \latin{et~al.}(1996)Perdew, Burke, and Ernzerhof]{Perdew1996}
Perdew,~J.~P.; Burke,~K.; Ernzerhof,~M. {Generalized gradient approximation
  made simple}. \emph{Physical Review Letters} \textbf{1996}, \emph{77},
  3865--3868\relax
\mciteBstWouldAddEndPuncttrue
\mciteSetBstMidEndSepPunct{\mcitedefaultmidpunct}
{\mcitedefaultendpunct}{\mcitedefaultseppunct}\relax
\EndOfBibitem
\bibitem[Grimme \latin{et~al.}(2010)Grimme, Antony, Ehrlich, and
  Krieg]{Grimme2010}
Grimme,~S.; Antony,~J.; Ehrlich,~S.; Krieg,~H. {A consistent and accurate ab
  initio parametrization of density functional dispersion correction (DFT-D)
  for the 94 elements H-Pu}. \emph{Journal of Chemical Physics} \textbf{2010},
  \emph{132}\relax
\mciteBstWouldAddEndPuncttrue
\mciteSetBstMidEndSepPunct{\mcitedefaultmidpunct}
{\mcitedefaultendpunct}{\mcitedefaultseppunct}\relax
\EndOfBibitem
\bibitem[Grimme \latin{et~al.}(2011)Grimme, Ehrlich, and Goerigk]{Grimme2011}
Grimme,~S.; Ehrlich,~S.; Goerigk,~L. {Effect of the Damping Function in
  Dispersion Corrected Density Functional Theory}. \emph{Journal of
  Computational Chemistry} \textbf{2011}, \emph{32}, 1456\relax
\mciteBstWouldAddEndPuncttrue
\mciteSetBstMidEndSepPunct{\mcitedefaultmidpunct}
{\mcitedefaultendpunct}{\mcitedefaultseppunct}\relax
\EndOfBibitem
\bibitem[Bl{\"{o}}chl(1994)]{Blochl1994}
Bl{\"{o}}chl,~P.~E. {Projector augmented-wave method}. \emph{Physical Review B}
  \textbf{1994}, \emph{50}, 17953--17979\relax
\mciteBstWouldAddEndPuncttrue
\mciteSetBstMidEndSepPunct{\mcitedefaultmidpunct}
{\mcitedefaultendpunct}{\mcitedefaultseppunct}\relax
\EndOfBibitem
\bibitem[Kresse and Joubert(1999)Kresse, and Joubert]{Kresse1999}
Kresse,~G.; Joubert,~D. {From ultrasoft pseudopotentials to the projector
  augmented-wave method}. \emph{Physical Review B - Condensed Matter and
  Materials Physics} \textbf{1999}, \emph{59}, 1758--1775\relax
\mciteBstWouldAddEndPuncttrue
\mciteSetBstMidEndSepPunct{\mcitedefaultmidpunct}
{\mcitedefaultendpunct}{\mcitedefaultseppunct}\relax
\EndOfBibitem
\bibitem[Plimpton(1995)]{Plimpton1997}
Plimpton,~S. {Fast Parallel Algorithms for Short-Range Molecular Dynamics}.
  \emph{Journal of Computational Physics} \textbf{1995}, \emph{117},
  1--19\relax
\mciteBstWouldAddEndPuncttrue
\mciteSetBstMidEndSepPunct{\mcitedefaultmidpunct}
{\mcitedefaultendpunct}{\mcitedefaultseppunct}\relax
\EndOfBibitem
\bibitem[Leven \latin{et~al.}(2014)Leven, Azuri, Kronik, and Hod]{Leven2014}
Leven,~I.; Azuri,~I.; Kronik,~L.; Hod,~O. {Inter-layer potential for hexagonal
  boron nitride}. \emph{Journal of Chemical Physics} \textbf{2014}, \emph{140},
  104106\relax
\mciteBstWouldAddEndPuncttrue
\mciteSetBstMidEndSepPunct{\mcitedefaultmidpunct}
{\mcitedefaultendpunct}{\mcitedefaultseppunct}\relax
\EndOfBibitem
\bibitem[Leven \latin{et~al.}(2016)Leven, Maaravi, Azuri, Kronik, and
  Hod]{Leven2016}
Leven,~I.; Maaravi,~T.; Azuri,~I.; Kronik,~L.; Hod,~O. {Interlayer Potential
  for Graphene/h-BN Heterostructures}. \emph{Journal of Chemical Theory and
  Computation} \textbf{2016}, \emph{12}, 2896--2905\relax
\mciteBstWouldAddEndPuncttrue
\mciteSetBstMidEndSepPunct{\mcitedefaultmidpunct}
{\mcitedefaultendpunct}{\mcitedefaultseppunct}\relax
\EndOfBibitem
\bibitem[Maaravi \latin{et~al.}(2017)Maaravi, Leven, Azuri, Kronik, and
  Hod]{Maaravi2017}
Maaravi,~T.; Leven,~I.; Azuri,~I.; Kronik,~L.; Hod,~O. {Interlayer Potential
  for Homogeneous Graphene and Hexagonal Boron Nitride Systems:
  Reparametrization for Many-Body Dispersion Effects}. \emph{Journal of
  Physical Chemistry C} \textbf{2017}, \emph{121}, 22826--22835\relax
\mciteBstWouldAddEndPuncttrue
\mciteSetBstMidEndSepPunct{\mcitedefaultmidpunct}
{\mcitedefaultendpunct}{\mcitedefaultseppunct}\relax
\EndOfBibitem
\bibitem[Solozhenko \latin{et~al.}(1995)Solozhenko, Will, and
  Elf]{Solozhenko1995}
Solozhenko,~V.~L.; Will,~G.; Elf,~F. {Isothermal compression of hexagonal
  graphite-like boron nitride up to 12 GPa}. \emph{Solid State Communications}
  \textbf{1995}, \emph{96}, 1--3\relax
\mciteBstWouldAddEndPuncttrue
\mciteSetBstMidEndSepPunct{\mcitedefaultmidpunct}
{\mcitedefaultendpunct}{\mcitedefaultseppunct}\relax
\EndOfBibitem
\bibitem[Cerda and Mahadevan(2003)Cerda, and Mahadevan]{Cerda2003}
Cerda,~E.; Mahadevan,~L. {Geometry and Physics of Wrinkling}. \emph{Physical
  Review Letters} \textbf{2003}, \emph{90}, 074302\relax
\mciteBstWouldAddEndPuncttrue
\mciteSetBstMidEndSepPunct{\mcitedefaultmidpunct}
{\mcitedefaultendpunct}{\mcitedefaultseppunct}\relax
\EndOfBibitem
\bibitem[Bosak \latin{et~al.}(2006)Bosak, Serrano, Krisch, Watanabe, Taniguchi,
  and Kanda]{Bosak2006}
Bosak,~A.; Serrano,~J.; Krisch,~M.; Watanabe,~K.; Taniguchi,~T.; Kanda,~H.
  {Elasticity of hexagonal boron nitride: Inelastic x-ray scattering
  measurements}. \emph{Physical Review B - Condensed Matter and Materials
  Physics} \textbf{2006}, \emph{73}, 041402(R)\relax
\mciteBstWouldAddEndPuncttrue
\mciteSetBstMidEndSepPunct{\mcitedefaultmidpunct}
{\mcitedefaultendpunct}{\mcitedefaultseppunct}\relax
\EndOfBibitem
\bibitem[Graziano \latin{et~al.}(2012)Graziano, Klime{\v{s}}, Fernandez-Alonso,
  and Michaelides]{Graziano2012}
Graziano,~G.; Klime{\v{s}},~J.; Fernandez-Alonso,~F.; Michaelides,~A. {Improved
  description of soft layered materials with van der Waals density functional
  theory}. \emph{Journal of Physics: Condensed Matter} \textbf{2012},
  \emph{24}, 424216\relax
\mciteBstWouldAddEndPuncttrue
\mciteSetBstMidEndSepPunct{\mcitedefaultmidpunct}
{\mcitedefaultendpunct}{\mcitedefaultseppunct}\relax
\EndOfBibitem
\bibitem[Ohba \latin{et~al.}(2001)Ohba, Miwa, Nagasako, and Fukumoto]{Ohba2001}
Ohba,~N.; Miwa,~K.; Nagasako,~N.; Fukumoto,~A. {First-principles study on
  structural, dielectric, and dynamical properties for three BN polytypes}.
  \emph{Physical Review B - Condensed Matter and Materials Physics}
  \textbf{2001}, \emph{63}, 115207\relax
\mciteBstWouldAddEndPuncttrue
\mciteSetBstMidEndSepPunct{\mcitedefaultmidpunct}
{\mcitedefaultendpunct}{\mcitedefaultseppunct}\relax
\EndOfBibitem
\bibitem[Murnaghan(1944)]{Murnaghan1944}
Murnaghan,~F.~D. {The Compressibility of Media Under Extreme Pressure}.
  \emph{Proceedings of the National Academy of Sciences of the United States of
  America} \textbf{1944}, \emph{30}, 244--247\relax
\mciteBstWouldAddEndPuncttrue
\mciteSetBstMidEndSepPunct{\mcitedefaultmidpunct}
{\mcitedefaultendpunct}{\mcitedefaultseppunct}\relax
\EndOfBibitem
\bibitem[Ma \latin{et~al.}(2016)Ma, Tocci, Michaelides, and Aeppli]{Ma2016}
Ma,~M.; Tocci,~G.; Michaelides,~A.; Aeppli,~G. {Fast diffusion of water
  nanodroplets on graphene}. \emph{Nature Materials} \textbf{2016}, \emph{15},
  66--71\relax
\mciteBstWouldAddEndPuncttrue
\mciteSetBstMidEndSepPunct{\mcitedefaultmidpunct}
{\mcitedefaultendpunct}{\mcitedefaultseppunct}\relax
\EndOfBibitem
\bibitem[Alf{\`{e}}(2009)]{Alfe2009}
Alf{\`{e}},~D. {PHON: A program to calculate phonons using the small
  displacement method}. \emph{Computer Physics Communications} \textbf{2009},
  \emph{180}, 2622--2633\relax
\mciteBstWouldAddEndPuncttrue
\mciteSetBstMidEndSepPunct{\mcitedefaultmidpunct}
{\mcitedefaultendpunct}{\mcitedefaultseppunct}\relax
\EndOfBibitem
\bibitem[Giannozzi \latin{et~al.}(1991)Giannozzi, de~Gironcoli, Pavone, and
  Baroni]{Giannozzi1991}
Giannozzi,~P.; de~Gironcoli,~S.; Pavone,~P.; Baroni,~S. {Ab initio calculation
  of phonon dispersions in semiconductors}. \emph{Physical Review B}
  \textbf{1991}, \emph{43}, 7231--7242\relax
\mciteBstWouldAddEndPuncttrue
\mciteSetBstMidEndSepPunct{\mcitedefaultmidpunct}
{\mcitedefaultendpunct}{\mcitedefaultseppunct}\relax
\EndOfBibitem
\bibitem[Gonze \latin{et~al.}(1994)Gonze, Charlier, Allan, and
  Teter]{Gonze1994}
Gonze,~X.; Charlier,~J.~C.; Allan,~D.~C.; Teter,~M.~P. {Interatomic force
  constants from first principles: The case of $\alpha$-quartz}. \emph{Physical
  Review B} \textbf{1994}, \emph{50}, 13035--13038\relax
\mciteBstWouldAddEndPuncttrue
\mciteSetBstMidEndSepPunct{\mcitedefaultmidpunct}
{\mcitedefaultendpunct}{\mcitedefaultseppunct}\relax
\EndOfBibitem
\bibitem[Gonze and Lee(1997)Gonze, and Lee]{Gonze1997}
Gonze,~X.; Lee,~C. {Dynamical matrices, Born effective charges, dielectric
  permittivity tensors, and interatomic force constants from density-functional
  perturbation theory}. \emph{Physical Review B - Condensed Matter and
  Materials Physics} \textbf{1997}, \emph{55}, 10355--10368\relax
\mciteBstWouldAddEndPuncttrue
\mciteSetBstMidEndSepPunct{\mcitedefaultmidpunct}
{\mcitedefaultendpunct}{\mcitedefaultseppunct}\relax
\EndOfBibitem
\bibitem[Baroni \latin{et~al.}(2001)Baroni, de~Gironcoli, and {Dal
  Corso}]{Baroni2001}
Baroni,~S.; de~Gironcoli,~S.; {Dal Corso},~A. {Phonons and related crystal
  properties from density-functional perturbation theory}. \emph{Reviews of
  Modern Physics} \textbf{2001}, \emph{73}, 515 -- 557\relax
\mciteBstWouldAddEndPuncttrue
\mciteSetBstMidEndSepPunct{\mcitedefaultmidpunct}
{\mcitedefaultendpunct}{\mcitedefaultseppunct}\relax
\EndOfBibitem
\bibitem[Sohier \latin{et~al.}(2017)Sohier, Gibertini, Calandra, Mauri, and
  Marzari]{Sohier2017}
Sohier,~T.; Gibertini,~M.; Calandra,~M.; Mauri,~F.; Marzari,~N. {Breakdown of
  Optical Phonons' Splitting in Two-Dimensional Materials}. \emph{Nano Letters}
  \textbf{2017}, \emph{17}, 3758--3763\relax
\mciteBstWouldAddEndPuncttrue
\mciteSetBstMidEndSepPunct{\mcitedefaultmidpunct}
{\mcitedefaultendpunct}{\mcitedefaultseppunct}\relax
\EndOfBibitem
\bibitem[Giannozzi \latin{et~al.}(2009)Giannozzi, Baroni, Bonini, Calandra,
  Car, Cavazzoni, Ceresoli, Chiarotti, Cococcioni, Dabo, {Dal Corso}, {De
  Gironcoli}, Fabris, Fratesi, Gebauer, Gerstmann, Gougoussis, Kokalj, Lazzeri,
  Martin-Samos, Marzari, Mauri, Mazzarello, Paolini, Pasquarello, Paulatto,
  Sbraccia, Scandolo, Sclauzero, Seitsonen, Smogunov, Umari, and
  Wentzcovitch]{Giannozzi2009}
Giannozzi,~P.; Baroni,~S.; Bonini,~N.; Calandra,~M.; Car,~R.; Cavazzoni,~C.;
  Ceresoli,~D.; Chiarotti,~G.~L.; Cococcioni,~M.; Dabo,~I. \latin{et~al.}
  {QUANTUM ESPRESSO: A modular and open-source software project for quantum
  simulations of materials}. \emph{Journal of Physics Condensed Matter}
  \textbf{2009}, \emph{21}, 395502\relax
\mciteBstWouldAddEndPuncttrue
\mciteSetBstMidEndSepPunct{\mcitedefaultmidpunct}
{\mcitedefaultendpunct}{\mcitedefaultseppunct}\relax
\EndOfBibitem
\bibitem[Giannozzi \latin{et~al.}(2017)Giannozzi, Andreussi, Brumme, Bunau,
  {Buongiorno Nardelli}, Calandra, Car, Cavazzoni, Ceresoli, Cococcioni,
  Colonna, Carnimeo, {Dal Corso}, {De Gironcoli}, Delugas, Distasio, Ferretti,
  Floris, Fratesi, Fugallo, Gebauer, Gerstmann, Giustino, Gorni, Jia, Kawamura,
  Ko, Kokalj, K{\"{u}}c{\"{u}}kbenli, Lazzeri, Marsili, Marzari, Mauri, Nguyen,
  Nguyen, Otero-De-La-Roza, Paulatto, Ponc{\'{e}}, Rocca, Sabatini, Santra,
  Schlipf, Seitsonen, Smogunov, Timrov, Thonhauser, Umari, Vast, Wu, and
  Baroni]{Giannozzi2017}
Giannozzi,~P.; Andreussi,~O.; Brumme,~T.; Bunau,~O.; {Buongiorno Nardelli},~M.;
  Calandra,~M.; Car,~R.; Cavazzoni,~C.; Ceresoli,~D.; Cococcioni,~M.
  \latin{et~al.}  {Advanced capabilities for materials modelling with Quantum
  ESPRESSO}. \emph{Journal of Physics Condensed Matter} \textbf{2017},
  \emph{29}, 465901\relax
\mciteBstWouldAddEndPuncttrue
\mciteSetBstMidEndSepPunct{\mcitedefaultmidpunct}
{\mcitedefaultendpunct}{\mcitedefaultseppunct}\relax
\EndOfBibitem
\bibitem[Togo and Tanaka(2015)Togo, and Tanaka]{Togo2015}
Togo,~A.; Tanaka,~I. {First principles phonon calculations in materials
  science}. \emph{Scripta Materialia} \textbf{2015}, \emph{108}, 1--5\relax
\mciteBstWouldAddEndPuncttrue
\mciteSetBstMidEndSepPunct{\mcitedefaultmidpunct}
{\mcitedefaultendpunct}{\mcitedefaultseppunct}\relax
\EndOfBibitem
\bibitem[Geick \latin{et~al.}(1966)Geick, Perry, and Rupprecht]{Geick1966}
Geick,~R.; Perry,~C.~H.; Rupprecht,~G. {Normal modes in hexagonal boron
  nitride}. \emph{Physical Review} \textbf{1966}, \emph{146}, 543--547\relax
\mciteBstWouldAddEndPuncttrue
\mciteSetBstMidEndSepPunct{\mcitedefaultmidpunct}
{\mcitedefaultendpunct}{\mcitedefaultseppunct}\relax
\EndOfBibitem
\bibitem[Nemanich \latin{et~al.}(1981)Nemanich, Solin, and
  Martin]{Nemanich1981}
Nemanich,~R.~J.; Solin,~S.~A.; Martin,~R.~M. {Light scattering study of boron
  nitride microcrystals}. \emph{Physical Review B} \textbf{1981}, \emph{23},
  6348--6356\relax
\mciteBstWouldAddEndPuncttrue
\mciteSetBstMidEndSepPunct{\mcitedefaultmidpunct}
{\mcitedefaultendpunct}{\mcitedefaultseppunct}\relax
\EndOfBibitem
\bibitem[Reich \latin{et~al.}(2005)Reich, Ferrari, Arenal, Loiseau, Bello, and
  Robertson]{Reich2005}
Reich,~S.; Ferrari,~A.~C.; Arenal,~R.; Loiseau,~A.; Bello,~I.; Robertson,~J.
  {Resonant Raman scattering in cubic and hexagonal boron nitride}.
  \emph{Physical Review B - Condensed Matter and Materials Physics}
  \textbf{2005}, \emph{71}, 205201\relax
\mciteBstWouldAddEndPuncttrue
\mciteSetBstMidEndSepPunct{\mcitedefaultmidpunct}
{\mcitedefaultendpunct}{\mcitedefaultseppunct}\relax
\EndOfBibitem
\bibitem[Serrano \latin{et~al.}(2007)Serrano, Bosak, Arenal, Krisch, Watanabe,
  Taniguchi, Kanda, Rubio, and Wirtz]{Serrano2007}
Serrano,~J.; Bosak,~A.; Arenal,~R.; Krisch,~M.; Watanabe,~K.; Taniguchi,~T.;
  Kanda,~H.; Rubio,~A.; Wirtz,~L. {Vibrational Properties of Hexagonal Boron
  Nitride : Inelastic X-Ray Scattering and Ab Initio Calculations}.
  \emph{Physical Review Letters} \textbf{2007}, \emph{98}, 095503\relax
\mciteBstWouldAddEndPuncttrue
\mciteSetBstMidEndSepPunct{\mcitedefaultmidpunct}
{\mcitedefaultendpunct}{\mcitedefaultseppunct}\relax
\EndOfBibitem
\bibitem[Wirtz \latin{et~al.}(2003)Wirtz, Rubio, {De La Concha}, and
  Loiseau]{Wirtz2003}
Wirtz,~L.; Rubio,~A.; {De La Concha},~R.; Loiseau,~A. {Ab initio calculations
  of the lattice dynamics of boron nitride nanotubes}. \emph{Physical Review B
  - Condensed Matter and Materials Physics} \textbf{2003}, \emph{68},
  045425\relax
\mciteBstWouldAddEndPuncttrue
\mciteSetBstMidEndSepPunct{\mcitedefaultmidpunct}
{\mcitedefaultendpunct}{\mcitedefaultseppunct}\relax
\EndOfBibitem
\bibitem[Zakharchenko \latin{et~al.}(2010)Zakharchenko, Los, Katsnelson, and
  Fasolino]{Zakharchenko2010}
Zakharchenko,~K.~V.; Los,~J.~H.; Katsnelson,~M.~I.; Fasolino,~A. {Atomistic
  simulations of structural and thermodynamic properties of bilayer graphene}.
  \emph{Physical Review B - Condensed Matter and Materials Physics}
  \textbf{2010}, \emph{81}, 235439\relax
\mciteBstWouldAddEndPuncttrue
\mciteSetBstMidEndSepPunct{\mcitedefaultmidpunct}
{\mcitedefaultendpunct}{\mcitedefaultseppunct}\relax
\EndOfBibitem
\bibitem[Slotman and Fasolino(2013)Slotman, and Fasolino]{Slotman2013}
Slotman,~G.~J.; Fasolino,~A. {Structure, stability and defects of single layer
  hexagonal BN in comparison to graphene}. \emph{Journal of Physics Condensed
  Matter} \textbf{2013}, \emph{25}, 045009\relax
\mciteBstWouldAddEndPuncttrue
\mciteSetBstMidEndSepPunct{\mcitedefaultmidpunct}
{\mcitedefaultendpunct}{\mcitedefaultseppunct}\relax
\EndOfBibitem
\bibitem[Katsnelson and Fasolino(2013)Katsnelson, and Fasolino]{Katsnelson2013}
Katsnelson,~M.~I.; Fasolino,~A. {Graphene as a prototype crystalline membrane}.
  \emph{Accounts of Chemical Research} \textbf{2013}, \emph{46}, 97--105\relax
\mciteBstWouldAddEndPuncttrue
\mciteSetBstMidEndSepPunct{\mcitedefaultmidpunct}
{\mcitedefaultendpunct}{\mcitedefaultseppunct}\relax
\EndOfBibitem
\bibitem[Xu \latin{et~al.}(2014)Xu, Neek-Amal, Barber, Schoelz, Ackerman,
  Thibado, Sadeghi, and Peeters]{Xu2014}
Xu,~P.; Neek-Amal,~M.; Barber,~S.~D.; Schoelz,~J.~K.; Ackerman,~M.~L.;
  Thibado,~P.~M.; Sadeghi,~A.; Peeters,~F.~M. {Unusual ultra-low-frequency
  fluctuations in freestanding graphene}. \emph{Nature Communications}
  \textbf{2014}, \emph{5}, 3720\relax
\mciteBstWouldAddEndPuncttrue
\mciteSetBstMidEndSepPunct{\mcitedefaultmidpunct}
{\mcitedefaultendpunct}{\mcitedefaultseppunct}\relax
\EndOfBibitem
\bibitem[Nelson and Peliti(1987)Nelson, and Peliti]{Nelson1987}
Nelson,~D.~R.; Peliti,~L. {Fluctuations in Membranes With Crystalline and
  Hexatic Order.} \emph{Journal de physique Paris} \textbf{1987}, \emph{48},
  1085--1092\relax
\mciteBstWouldAddEndPuncttrue
\mciteSetBstMidEndSepPunct{\mcitedefaultmidpunct}
{\mcitedefaultendpunct}{\mcitedefaultseppunct}\relax
\EndOfBibitem
\bibitem[{Le Doussal} and Radzihovsky(1992){Le Doussal}, and
  Radzihovsky]{LeDoussal1992}
{Le Doussal},~P.; Radzihovsky,~L. {Self-consistent theory of polymerized
  membranes}. \emph{Physical Review Letters} \textbf{1992}, \emph{69},
  1209--1212\relax
\mciteBstWouldAddEndPuncttrue
\mciteSetBstMidEndSepPunct{\mcitedefaultmidpunct}
{\mcitedefaultendpunct}{\mcitedefaultseppunct}\relax
\EndOfBibitem
\bibitem[Nelson \latin{et~al.}(2004)Nelson, Piran, and Weinberg]{Nelson2004}
Nelson,~D.~R., Piran,~T., Weinberg,~S., Eds. \emph{{Statistical Mechanics of
  Membranes and Surfaces}}, 2nd ed.; World Scientific: Singapore, 2004\relax
\mciteBstWouldAddEndPuncttrue
\mciteSetBstMidEndSepPunct{\mcitedefaultmidpunct}
{\mcitedefaultendpunct}{\mcitedefaultseppunct}\relax
\EndOfBibitem
\bibitem[Kownacki and Mouhanna(2009)Kownacki, and Mouhanna]{Kownacki2009}
Kownacki,~J.~P.; Mouhanna,~D. {Crumpling transition and flat phase of
  polymerized phantom membranes}. \emph{Physical Review E - Statistical,
  Nonlinear, and Soft Matter Physics} \textbf{2009}, \emph{79}, 040101(R)\relax
\mciteBstWouldAddEndPuncttrue
\mciteSetBstMidEndSepPunct{\mcitedefaultmidpunct}
{\mcitedefaultendpunct}{\mcitedefaultseppunct}\relax
\EndOfBibitem
\bibitem[Karssemeijer and Fasolino(2011)Karssemeijer, and
  Fasolino]{Karssemeijer2011}
Karssemeijer,~L.~J.; Fasolino,~A. {Phonons of graphene and graphitic materials
  derived from the empirical potential LCBOPII}. \emph{Surface Science}
  \textbf{2011}, \emph{605}, 1611--1615\relax
\mciteBstWouldAddEndPuncttrue
\mciteSetBstMidEndSepPunct{\mcitedefaultmidpunct}
{\mcitedefaultendpunct}{\mcitedefaultseppunct}\relax
\EndOfBibitem
\bibitem[Wu \latin{et~al.}(2013)Wu, Wang, Wei, Yang, and Dresselhaus]{Wu2013}
Wu,~J.; Wang,~B.; Wei,~Y.; Yang,~R.; Dresselhaus,~M. {Mechanics and
  mechanically tunable band gap in single-layer hexagonal boron-nitride}.
  \emph{Materials Research Letters} \textbf{2013}, \emph{1}, 200--206\relax
\mciteBstWouldAddEndPuncttrue
\mciteSetBstMidEndSepPunct{\mcitedefaultmidpunct}
{\mcitedefaultendpunct}{\mcitedefaultseppunct}\relax
\EndOfBibitem
\bibitem[Wei \latin{et~al.}(2013)Wei, Wang, Wu, Yang, and Dunn]{Wei2013}
Wei,~Y.; Wang,~B.; Wu,~J.; Yang,~R.; Dunn,~M.~L. {Bending rigidity and Gaussian
  bending stiffness of single-layered graphene}. \emph{Nano Letters}
  \textbf{2013}, \emph{13}, 26--30\relax
\mciteBstWouldAddEndPuncttrue
\mciteSetBstMidEndSepPunct{\mcitedefaultmidpunct}
{\mcitedefaultendpunct}{\mcitedefaultseppunct}\relax
\EndOfBibitem
\bibitem[Zakharchenko \latin{et~al.}(2009)Zakharchenko, Katsnelson, and
  Fasolino]{Zakharchenko2009}
Zakharchenko,~K.~V.; Katsnelson,~M.~I.; Fasolino,~A. {Finite temperature
  lattice properties of graphene beyond the quasiharmonic approximation}.
  \emph{Physical Review Letters} \textbf{2009}, \emph{102}, 046808\relax
\mciteBstWouldAddEndPuncttrue
\mciteSetBstMidEndSepPunct{\mcitedefaultmidpunct}
{\mcitedefaultendpunct}{\mcitedefaultseppunct}\relax
\EndOfBibitem
\bibitem[Thomas \latin{et~al.}(2017)Thomas, Ajith, and Valsakumar]{Thomas2017}
Thomas,~S.; Ajith,~K.~M.; Valsakumar,~M.~C. {Effect of ripples on the finite
  temperature elastic properties of hexagonal boron nitride using
  strain-fluctuation method}. \emph{Superlattices and Microstructures}
  \textbf{2017}, \emph{111}, 360--372\relax
\mciteBstWouldAddEndPuncttrue
\mciteSetBstMidEndSepPunct{\mcitedefaultmidpunct}
{\mcitedefaultendpunct}{\mcitedefaultseppunct}\relax
\EndOfBibitem
\bibitem[{De Lima} \latin{et~al.}(2015){De Lima}, M{\"{u}}ssnich, Manhabosco,
  Chacham, Batista, and {De Oliveira}]{DeLima2015}
{De Lima},~A.~L.; M{\"{u}}ssnich,~L.~A.; Manhabosco,~T.~M.; Chacham,~H.;
  Batista,~R.~J.; {De Oliveira},~A.~B. {Soliton instability and fold formation
  in laterally compressed graphene}. \emph{Nanotechnology} \textbf{2015},
  \emph{26}, 045707\relax
\mciteBstWouldAddEndPuncttrue
\mciteSetBstMidEndSepPunct{\mcitedefaultmidpunct}
{\mcitedefaultendpunct}{\mcitedefaultseppunct}\relax
\EndOfBibitem
\bibitem[Shankar \latin{et~al.}(2019)Shankar, Marchesini, and
  Petit]{Shankar2019}
Shankar,~R.; Marchesini,~S.; Petit,~C. {Enhanced Hydrolytic Stability of Porous
  Boron Nitride via the Control of Crystallinity, Porosity, and Chemical
  Composition}. \emph{Journal of Physical Chemistry C} \textbf{2019},
  \emph{123}, 4282--4290\relax
\mciteBstWouldAddEndPuncttrue
\mciteSetBstMidEndSepPunct{\mcitedefaultmidpunct}
{\mcitedefaultendpunct}{\mcitedefaultseppunct}\relax
\EndOfBibitem
\bibitem[Kroes \latin{et~al.}(2017)Kroes, Fasolino, and Katsnelson]{Kroes2017}
Kroes,~J.~M.; Fasolino,~A.; Katsnelson,~M.~I. {Density functional based
  simulations of proton permeation of graphene and hexagonal boron nitride}.
  \emph{Physical Chemistry Chemical Physics} \textbf{2017}, \emph{19},
  5813--5817\relax
\mciteBstWouldAddEndPuncttrue
\mciteSetBstMidEndSepPunct{\mcitedefaultmidpunct}
{\mcitedefaultendpunct}{\mcitedefaultseppunct}\relax
\EndOfBibitem
\bibitem[Liang \latin{et~al.}(2017)Liang, He, Wu, Ren, Guo, Kong, Iwai, Fujita,
  Gao, Guo, Liu, and Xu]{Liang2017}
Liang,~T.; He,~G.; Wu,~X.; Ren,~J.; Guo,~H.; Kong,~Y.; Iwai,~H.; Fujita,~D.;
  Gao,~H.; Guo,~H. \latin{et~al.}  {Permeation through graphene ripples}.
  \emph{2D Materials} \textbf{2017}, \emph{4}, 025010\relax
\mciteBstWouldAddEndPuncttrue
\mciteSetBstMidEndSepPunct{\mcitedefaultmidpunct}
{\mcitedefaultendpunct}{\mcitedefaultseppunct}\relax
\EndOfBibitem
\bibitem[Sun \latin{et~al.}(2020)Sun, Yang, Kuang, Stebunov, Xiong, Yu, Nair,
  Katsnelson, Yuan, Grigorieva, Lozada-Hidalgo, Wang, and Geim]{Sun2020}
Sun,~P.~Z.; Yang,~Q.; Kuang,~W.~J.; Stebunov,~Y.~V.; Xiong,~W.~Q.; Yu,~J.;
  Nair,~R.~R.; Katsnelson,~M.~I.; Yuan,~S.~J.; Grigorieva,~I.~V. \latin{et~al.}
   {Limits on gas impermeability of graphene}. \emph{Nature} \textbf{2020},
  \emph{579}, 229--232\relax
\mciteBstWouldAddEndPuncttrue
\mciteSetBstMidEndSepPunct{\mcitedefaultmidpunct}
{\mcitedefaultendpunct}{\mcitedefaultseppunct}\relax
\EndOfBibitem
\end{mcitethebibliography}
  \providecommand{\latin}[1]{#1}
\makeatletter
\providecommand{\doi}
  {\begingroup\let\do\@makeother\dospecials
  \catcode`\{=1 \catcode`\}=2 \doi@aux}
\providecommand{\doi@aux}[1]{\endgroup\texttt{#1}}
\makeatother
\providecommand*\mcitethebibliography{\thebibliography}
\csname @ifundefined\endcsname{endmcitethebibliography}
  {\let\endmcitethebibliography\endthebibliography}{}


\end{document}